\documentclass[showpacs,preprint,aps,altaffilletter, altaffillsymbol]{revtex4}
\usepackage{graphicx}
\usepackage{amsmath}
\usepackage{float}

\usepackage{epsfig}
\begin{document}
\title{Quantum resonant effects in the delta-kicked rotor revisited}
\author{A. Ullah}
\affiliation{Department of Physics, University of Auckland, Private Bag 92019, Auckland, New Zealand}
\author{S. K. Ruddell}
\affiliation{Department of Physics, University of Auckland, Private Bag 92019, Auckland, New Zealand}
\author{J-A. Currivan}
\altaffiliation[Current address: ]{Massachusetts Institute of technology 77 MA, Bldg 13-3061, Cambridge, MA 02139.}
\affiliation{Department of Physics, University of Auckland, Private Bag 92019,
Auckland, New Zealand}
\author{M. D. Hoogerland}
\affiliation{Department of Physics, University of Auckland, Private Bag 92019, Auckland, New Zealand}
\newcommand{\kbar}{\mathchar'26\mkern-9muk}

\begin{abstract}
We review the theoretical model and experimental realization of the atom optics $\delta-$kicked rotor (AOKR), a paradigm of classical and quantum chaos. We have performed a number of experiments
with an all-optical Bose-Einstein condensate (BEC) in a periodic standing wave potential in an AOKR system. We discuss results of the investigation of the phenomena of quantum resonances in the
AOKR. An interesting feature of the momentum distribution of the atoms obtained as a result of short pulses of light, is the variance of the momentum distribution or the kinetic energy $\langle
p^{2}\rangle/2m$ in units of the recoil energy $E_{rec} = \hbar \omega_{rec}$. The energy of the system is examined as a function of pulse period for a range of kicks that allow the observation
of quantum resonances. In particular we study the behavior of these resonances for a large number of kicks. Higher order quantum resonant effects corresponding to the fractional Talbot time of
(1/4)$T_{T}$ and (1/5)$T_{T}$ for five and ten kicks have been observed. Moreover, we describe the effect of the initial momentum of the atoms on quantum resonances in the AOKR.
\end{abstract} 

\maketitle

\section{Introduction}
%
In classical dynamics, small perturbations in the initial conditions of a system may grow exponentially with time and produce large ones in the final states. This makes the motion practically
unpredictable, and the phenomenon is known as classical chaos ~\cite{Lichetenberg1992,Lieberman1972a,Kornfeld1982,Chirikov1979}.
On the other hand, the investigation of the behavior of quantum systems, the classical limit of which are chaotic, has brought up a new discipline in physics, which is now known as ``quantum
chaos''. The pioneering work by Giulio Casati and co-workers predicted that a quantum particle shows diffusion following the classical evolution, after a certain time known as the quantum break
time ~\cite{Casati1979,Chirikov1981,D.L.Shepelyanski1983}. The diffusion stops due to quantum interference, beyond the quantum break time, leading to dynamical localization
~\cite{Moore1995,Ammann1998} which is analogous to Anderson localization in solid state physics.

The strange connection between classical chaos and quantum chaos has forced
people to develop models that can describe systems that are chaotic. The kicked
rotor is one of the simplest models
that is useful to understand classical and quantum chaos. The classical model of
the kicked rotor is a simple pendulum that is kicked periodically by an external
force. The dynamics of the the
pendulum that exhibits chaos can be described by the Hamiltonian mechanics.
Classically a kicked rotor is a well known system that makes its quantum
analogue the best model to study quantum
chaos ~\cite{Moore1995,Ammann1998,Stockmann1999,Fishman1993}.

The quantum analogue of the classical kicked rotor is obtained by a cloud of ultra cold atoms interacting with a pulsed standing wave of laser light and is known as atom-optics kicked rotor
(AOKR). The atom-optics realization of the kicked rotor was first demonstrated by the Raizen group ~\cite{Moore1995} in 1995, and has been studied extensively theoretically and experimentally in
~\cite{Casati1979,Izrailev1980,Izrailev1980a,Saunders2007,Halkyard2008, Saunders2009,McDowall2009,Saif2005} and ~\cite{Wimberger2003,Wimberger2005,Deng1999,Oberthaler1999,Szriftgiser2002,
Duffy2004,Jones2007,Wayper2007,Currivan2009,Ullah2011} and the references therein. The AOKR model has allowed experimental studies of the phenomena of ``quantum resonances''
~\cite{Fishman1982,Wimberger2003} that occur due to the matter-wave Talbot effect ~\cite{Ryu2006,Deng1999,Ovchinnikov1999,Lepers2008}, analogous to the optical Talbot effect ~\cite{Talbot1836}.


An interesting feature of
the momentum distribution of the atoms obtained as a result of short pulses of
light, is the variance of the momentum distribution or the kinetic energy
$\langle p^{2}\rangle/2m$, in units of
recoil energy $E_{rec} = \hbar \omega_{rec}=\hbar^2 k_{L}^2/2m$, where
$k_L=2\pi/\lambda$, where $\lambda$ is the wavelength of the laser beam and $m$
the mass of the atom. The energy of the system is examined as a
function of pulse period.

In this paper, we present some of the known results along with our new results to illustrate the phenomena of quantum resonances in the AOKR. We also present experimental observation of higher
order quantum resonances corresponding to the fractional Talbot time for five and ten kicks. The results are in good agreement with those presented in Ryu {\em et. al.} ~\cite{Ryu2006}, where
higher order resonances were found in a similar experiment with a condensate of Na atoms. Moreover, in section III, the dependence of the initial momentum of the atoms on quantum resonances have
been discussed in detail. The results presented are in good agreement with those proposed in ~\cite{Saunders2007}.

\section{Theory}
The general Hamiltonian that describes the dynamics of atoms in the Atom-Optics kicked rotor experiment
can be expressed as
\begin{equation}
\ H=\frac{p^{2}}{2m}+\hbar\phi_{d}\cos(2k_{L}x) \displaystyle\sum\limits_{n=1}^N
\delta(t-nT),
\end{equation}
where $T$ is the kick period and $\phi_{d}$ is the kick strength. The kick strength $\phi_{d}=\Omega_{\rm eff} \tau$, which is determined by the effective Rabi frequency during the laser pulse
$\Omega_{\rm eff}$ and the pulse length $\tau$. The effective Rabi frequency is defined as $\Omega_{\rm eff}=\Omega^2/\Delta$.

\subsection{Evolution of the wave packet}
The time evolution operator that determines the evolution of the wave function
from one kick to immediately after the next kick is known as
the Floquet operator $\mathcal F$
~\cite{Currivan2009,Saunders2009a,Sadgrove2005}.
In general, the state of the system $|\psi(t_{n})\rangle$ after $n$ number of
kicks can be found by the Floquet operator acting on the initial
state $|\psi(t_{0})\rangle$ i.e.,
\begin{equation}
\ |\psi(t_{n})\rangle=\mathcal{F}^{n}|\psi(t_{0})\rangle.
\end{equation}
In the first part of the Hamiltonian describing the quantum kicked rotor, the system evolves freely via the operator
\begin{equation}
\ \hat{U}_{free}=e^{-i p^{2}T/2m\hbar},
\end{equation}
Then delta kicks act for a small period of time, and the evolution is given by the operator
\begin{equation}
\ \hat{U}_{kick}=e^{-i\phi_{d}\cos(2k_{L}x)\tau/\hbar}.
\end{equation}
Combining the above two operators, the Floquet operator is given by
\begin{equation}
\ \hat{\mathcal{F}}=  \underset{\hat{U}_{free}}{\underbrace{e^{-i p^{2}T/2m\hbar}}} \underset{\hat{U}_{kick}}{\underbrace{e^{-i \phi_{d}\cos(2k_{L}x)\tau/\hbar}}},
\end{equation}
or in a more general form as
\begin{equation}
\ \hat{\mathcal{F}}=\hat{U}_{free}\hat{U}_{kick}.
\end{equation}
From the definition above, the action of each kick on the atoms, which is
governed by $\hat{U}_{kick}$ is followed by a phase of free evolution
$\hat{U}_{free}$. A more detailed description of
the evolution of the quantum wave packet for a delta kicked rotor system can be
found in ~\cite{Saunders2007}.
\subsection{Simulating the AOKR}
In order to simulate the AOKR, the well-known split operator method for the time evolution of the Floquet operator has been used, which is ideally suitable for the experiment with $2^{16}$ grid
points, as described in detail in ~\cite{Currivan2009} and ~\cite{Ullah2011}. As described in the previous section, the time evolution of the quantum $\delta-$kicked rotor can be separated into
two parts: The evolution of the system during the time when laser pulses are applied on the atoms as kicks, and the evolution during the time of free expansion in between the kicks. The kick
potential is a diagonal operator in position space, whereas the free evolution is a diagonal operator in momentum space. Therefore, for the time evolution of the system, a Fourier transformation
is used in order to transform the wave function from position space into momentum space and back
again. 
The initial wave function representing the state of the system can be defined as
\begin{equation}
\ \psi(x)= \frac{1}{\sqrt{2\pi \sigma_{w}}} e^{-x^{2}/2 \sigma_{w}^{2}} e^{-i k_{i}x}.
\end{equation}
The width of the initial wave packet is chosen such that it covers several
wavelengths of the kick laser.

For the kicking part of the time evolution, the delta-function
is represented by a block function with a width $\tau\ll T$ and an area $\phi_{d}$ ~\cite{Currivan2009}.
The initial wave function $\psi(x)$  in position
space in the simulation is multiplied by
\begin{equation}
\exp{\{\frac{-i}\phi_d \}}.
\end{equation}


The wave function is then transformed from
position space into momentum space
represented by $\varphi(x)$ using an inverse Fourier transform. The free
evolution part of the system can be determined by multiplying the wave function
in momentum space by
\begin{equation}
\ \exp{\left\{\frac{-i}{\hbar} \frac{p^{2}}{2m} T_{free}\right\}},
\end{equation}
 which gives the final state in the momentum space $\varphi(x)$. The parameter
$T_{free}$ represents the pulse period in between the kicks. After the free
evolution part, the wave function
 is transformed back to position space by applying the Fourier transform and
gives the updated state $\psi(x)$. The process is repeated for each kick, giving
us the updated wave function each time
 after the evolution. The momentum and energy of the atoms after each kick is
then determined.
 \begin{figure}[!t]
   \center
\includegraphics[width=.9\columnwidth]{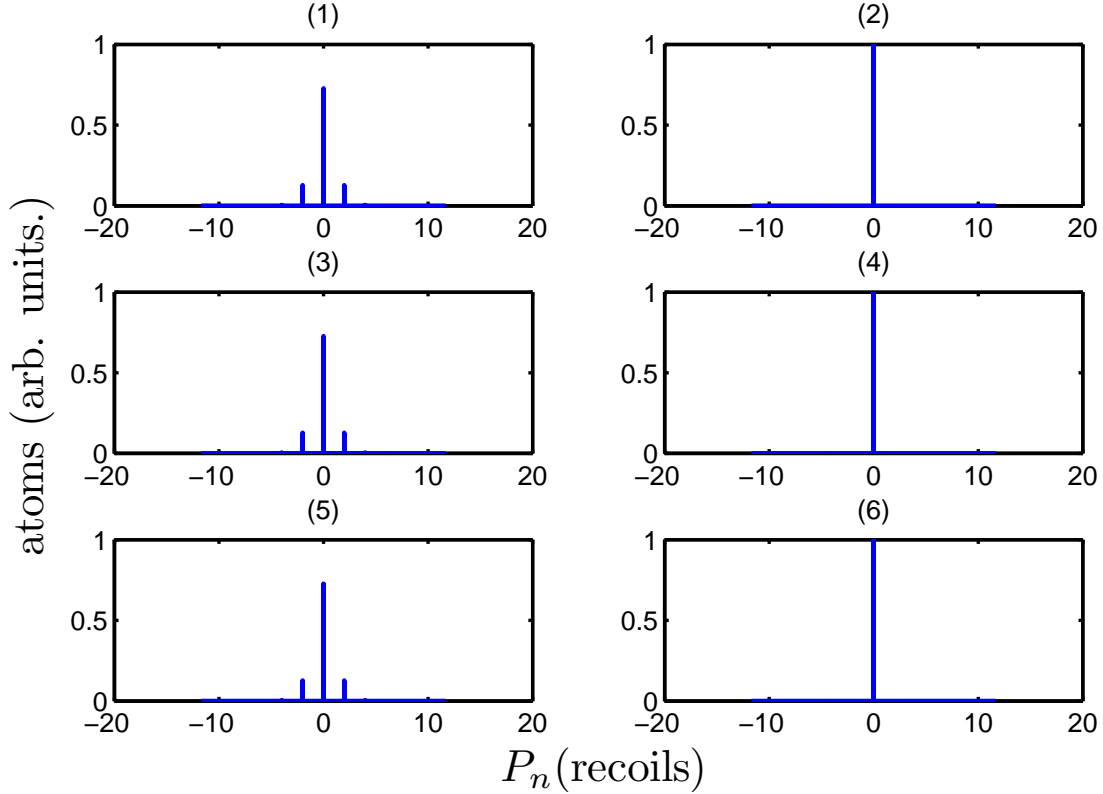}
   \caption{The simulated momentum distribution of the wave packet for the
quantum resonance case, i.e, $\kbar=4\pi$ is illustrated for each kick, with an initial $p=0$ momentum state.
   The distributions are shown for $N=1-6$ kicks, with a kick strength of
$\phi_{d} \approx 1.5$.}
   \label{reskicks}
\end{figure}
\subsection{Quantum resonances in AOKR}
Quantum resonances are well known quantum effects that occur for specific initial conditions and certain values of the kicking period. We investigate these resonances in the AOKR by considering
the atoms diffracted by a standing wave potential. A ``quantum resonance'' exists for $T=mT_{T}$, $T_{T}$ being the Talbot time, where all kicks add coherently and energy grows quadratically
with the number of kicks. An ``anti-resonance'' is observed at $T=(m+\frac{1}{2})\pi/2\omega_{rec}$, where the effect of each kick is effectively negated by the following kick.

\begin{figure}[!t]
   \center
   \includegraphics[width=.9\columnwidth]{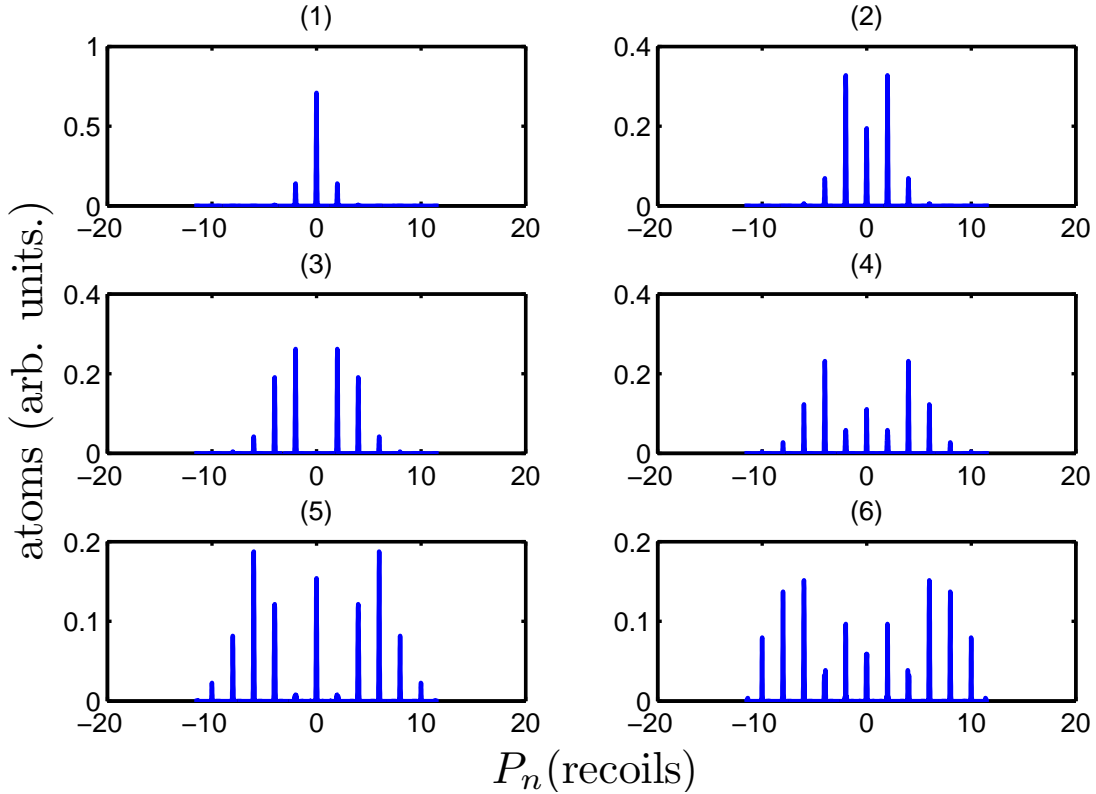}
   \caption{The simulated momentum distribution of the wave packet after kicks
for the quantum anti-resonance case, i.e., $\kbar=2\pi$ is shown for each kick. The parameters are $N=1-6$, with
   a kick strength of $\phi_{d} \approx 1.5$.}
   \label{antireskicks}
\end{figure}

The free evolution term is of primary importance in describing the quantum kicked rotor. If the initial state is assumed to be a momentum eigenstate $|m\rangle$, and let the operator
$\hat{U}_{free}$ act on it such that $\hat{\rho}|m\rangle=m\kbar|m\rangle$, then the evolution is different for different values of $\kbar$. If $\kbar$ is an even multiple of $2\pi$, then the
free evolution results in unity. This means that the free time evolution in this case does not have any effect on the state vector of the system. The system responds only to the $\delta-$kicks
~\cite{Saunders2009a}, and the evolution is fully governed during interaction with the kicking potential. This is the condition for quantum resonance ~\cite{Izrailev1980}. On the other hand, if
$\kbar$ is an odd multiple of $2\pi$, then the initial momentum eigenstate acquires different quantum phases for even and odd $m$. The state after the free evolution acquires a quantum phase of
$+1$ for even values of $m$, whereas for odd $m$ the quantum phase results in $-1$. For odd values of the momentum components of the wave-function, the phase of $-1$ leads to an oscillation of
the energy of the system. Therefore, $\kbar=2\pi$ is known as the quantum anti-resonance condition for the kicked rotor ~\cite{Oskay2000,Sadgrove2005a}. The phenomenon of quantum resonance and
quantum anti-resonance is demonstrated in the simulations in Figures~\ref{reskicks} and~\ref{antireskicks} respectively.

The general expression for the $q_{th}$ order momentum moment of the
$\delta-$kicked particle after time $nT$ as expressed in ~\cite{Saunders2007} is
given by
\begin{equation}
{\left\langle {p}^{q}\right\rangle}_{n}=(\hbar k)^{q}
\sum\limits_{j=-\infty}^\infty
\left[J_{j-k}\left(\phi_{d}\frac{\sin(n\Upsilon)}{\sin\Upsilon}\right)\right]
^2(j+\beta)^q.
\end{equation}
The parameter $\Upsilon$ is defined as
\begin{equation}
\Upsilon= \frac{1}{2}\pi (1+2\beta)l,
\end{equation}
where $\beta$ is the quasimomentum. Quantum resonant and antiresonant behavior of the system can be characterized by the second order momentum moment for $q=2$, and is proportional to the
kinetic energy.

%

As discussed in the previous sections, the evolution of the system can be
determined by the time evolution operator. In the case of quantum resonance, the
free evolution part is unity, therefore
for $N$ kicks the evolution operator is written as
\begin{equation}
\hat{U}^N= e^{\left(i\frac{\kappa}{\kbar}N\cos(2k_{L}x)\right)}
\end{equation}
where $\kappa/\kbar=\phi_{d}$ is the kick strength. The final state of the
system is as if a single $\delta-$kick had been applied with a strength $N$
times the original kick strength
$\phi_{d}$. The probability for an initial momentum state $|m\rangle$, to be
populated in the final state $|n\rangle$ after $N$ kicks is given by
\begin{align}
P_{n} & = \left|\left<n|\hat{U^{N}}|m\right>\right|^2 \\
      & =
\left|\left<n|e^{\left(i\frac{\kappa}{\kbar}N\cos(2k_{L}x)\right)} |m\right>\right|^{2}
\end{align}
Using the definition of orthogonality of the momentum eigenstate $\left <n|m\right >=\delta_{nm}$ and the identity $e^{(i k\sin x)}= \sum_{s}J_{s}e^{i sx }$, where $J$ is the ordinary Bessel
function of order $s$, we get
\begin{equation}
P_{n}  =  J^{2}_{n-m}\left(\frac{\kappa N}{\kbar}\right).
\end{equation}
%
For a system that starts in a zero momentum eigenstate, the final momentum
distribution becomes
\begin{equation}
P_{n}  =  J^{2}_{n}\left(\frac{\kappa N}{\kbar}\right).
\end{equation}
The energy of the ensemble after $N$ kicks can also be found at the quantum
resonance, and is given by
\begin{equation}
E_{n} = \sum_{n} n^{2} P_{n} = \frac{1}{2}\frac{\kappa^{2} N^{2}}{\kbar^2}.
\end{equation}
For a kicked rotor starting in the zero momentum state, the above result shows a
quadratic growth in energy for $N$ kicks for the quantum resonance.
\subsection{Experimental effects}
In the experiment, the initial momentum distribution from the BEC has a width of $\sigma \sim 0.18$ recoils. The initial width associated with the BEC qualitatively alters the momentum
distribution after the kick sequence. In an ideal case the initial momentum distribution is a delta function at $p_{i}$, however in reality, the initial momentum distribution has a width.
Because of this width, there are some atoms at a momentum $p_{i} + \Delta_{p}$, where the quantity $\Delta_{p}$ is much smaller than one recoil. In order to take into account this initial
momentum width, we performed the simulation for a range of initial momenta. In Fig.~\ref{deltap}, the probability of finding the atom at a final momentum is plotted for a range of initial
momenta for both antiresonance (top) and resonance (bottom). The initial momentum distribution of the BEC is also displayed (as a solid curve). The effect of the initial momentum on the system
can be observed as a transfer of the probability of atoms to states at small offset momenta. As the probability is very sensitive to the initial momentum, this sensitivity therefore translates
into peaks appearing at a small offset momentum $p_{i} + \Delta_{p}$ which are not present at $p_{i}$, where $p_{i}$ is equal to one recoil in this case. Conversely, this can also be seen as a
peak disappearing at $p_{i} + \Delta_{p}$ which was present at $p_{i}$.
The resulting momentum distributions are then
added incoherently, weighed by the height of the momentum distribution of the original BEC at $\Delta_{p}$. The final momentum distribution is obtained by the sum of horizontal profiles in
Fig.~\ref{deltap} weighed by the initial momentum distribution.

%
\section{Experimental observation of quantum resonances}
\subsection{Experimental Setup}
\label{sec:expt}
\begin{figure}[!t]
   \center
   \includegraphics[width=.7\columnwidth]{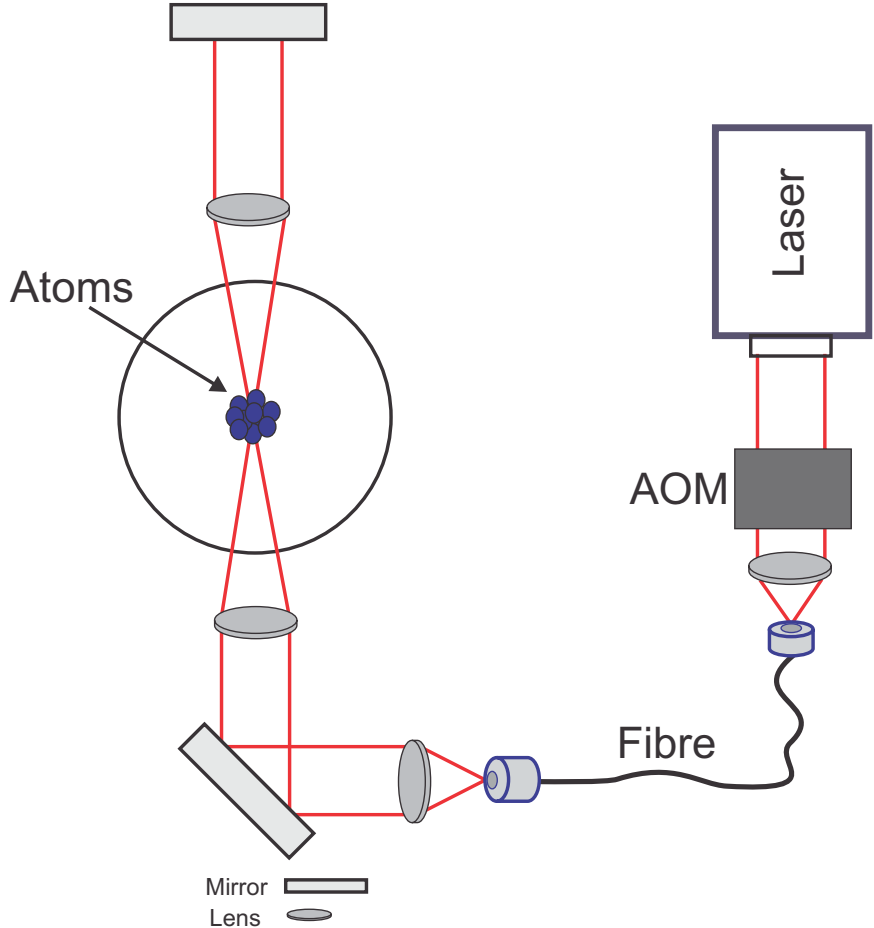}
   \caption{A sketch of the kick laser setup. A linearly polarized laser beam is
retro-reflected to form a standing wave potential. The laser is stabilized to the $F=2\rightarrow F'=3$ transition in $^{85}$Rb.}
   \label{stwaves}
\end{figure}
We use a double magneto-optical trap (MOT) configuration for the formation of all-optical Bose-Einstein Condensate (BEC) of ($F=1$) $^{87}$Rb at a temperature of 50~nK. The all-optical BEC of
$\sim$ $2.10^{4}$ atoms is formed in a cross dipole trap created by a pair of intersecting focused CO$_2$ laser beams. A detailed description of the experimental setup can be found in
~\cite{Wenas2008}.

We realise the AOKR by pulsing a near resonant optical standing wave, derived from a 780~nm diode laser, onto a BEC of $^{87}$Rb atoms. The kick laser is locked to the $S_{\frac{1}{2}},
F=2\rightarrow P_{\frac{3}{2}}, F'=3$ transition in the $^{85}$Rb isotope. Hence, the laser frequency is detuned by 2.45~GHz from the relevant $F=1\rightarrow F=2$ resonance frequency. The laser
beam from the diode laser passes through an acousto-optic modulator (AOM) for fast switching. After passing through the AOM, the beam passes through a single mode optical fibre and is focused
onto the BEC. The beam diameter at the focus is $\sim$ 100 $\mu$m, much larger than the size of the BEC ($\sim$10 $\mu$m), thereby yielding a constant interaction strength over the BEC, while
increasing the intensity.  After passing through the center of the trap the linearly polarized beam is then retro-reflected to produce a standing wave. The kick laser setup is shown in
Fig.~\ref{stwaves}.

 In the AOKR experiment, the BEC is illuminated with short pulses of light in the form
of a standing wave. The standing wave potential or the interaction potential
acts like a diffraction grating, which
changes the atomic momentum, and the BEC eventually splits into a number of
momentum components. An alternate picture is that the atoms absorb a
photon of momentum $\hbar k$ from one beam, and then provide another photon
through stimulated emission in the other beam. The diffracted components of
momentum are therefore separated by $\pm
2\hbar k_{L}$.
In order to reduce mean field effects, the trap that contains the BEC is turned off $500$ $\mu$s before the kick sequence is applied. A shutter is used in the laser beam to ensure that it is
totally extinguished during the evaporation phase to create the BEC.

The periodic potential is related directly to the intensity of the laser beam and inversely to its detuning from the atomic resonance. High laser power allows the experiment to be done at large
detunings to suppress spontaneous emission. Therefore, for the interaction potential it is desirable to get as much power as possible from the kick laser. The total laser power of the experiment
can be varied up to 2 mW. The laser pulses are so short that the evolution of the wave packet due to its momentum can be ignored during the kicking time. Atoms satisfying this condition are said
to be in the Raman-Nath regime ~\cite{Currivan2009,Sadgrove2005}.
\begin{figure}[!t]
   \center
   \includegraphics[width=.9\columnwidth]{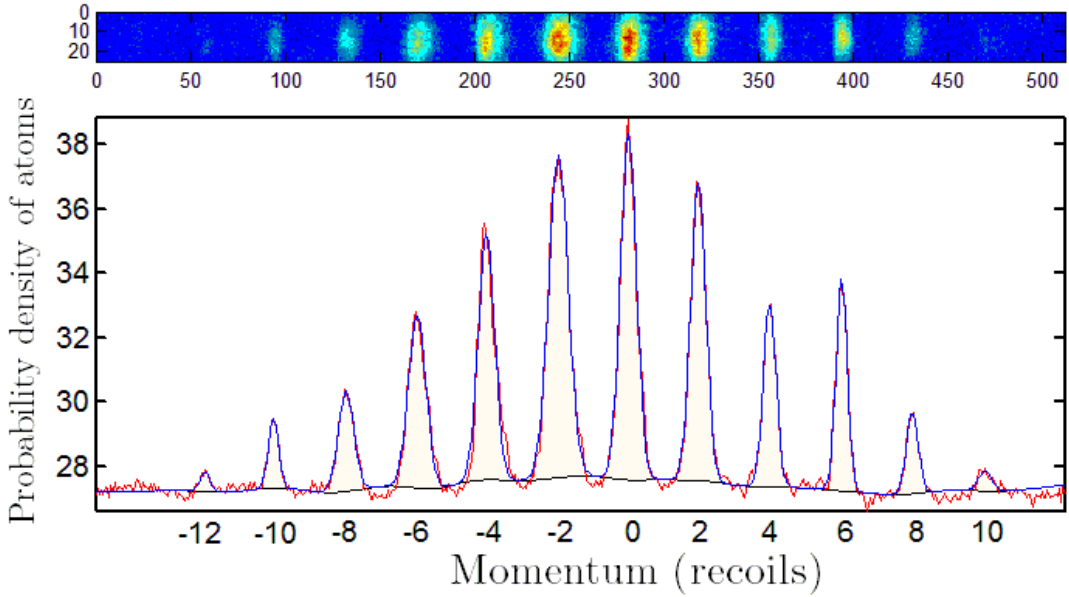}
   \caption{The momentum distribution of the atoms fitted by a Gaussian
distribution to obtain the energy.}
   \label{fitgauss}
\end{figure}

The momentum distribution of the atoms after the kick sequence is measured by absorption imaging using the time-of-flight technique, with a typical flight time of 5 to 10 ms.  Just prior to
imaging, the atoms are optically pumped to the $F=2$ state by a 100 $\mu$s pulse on the $F=1\rightarrow F'=2$ repump transition. An absorption image is then obtained using a probe laser which is
tuned to the $S_{\frac{1}{2}}, F=2\rightarrow P_{\frac{3}{2}}, F'=3$ transition. The two dimensional momentum distributions obtained are summed over the width of the cloud to obtain a one
dimensional momentum distribution. The energy of the atoms is then obtained by the variance of the momentum distribution by fitting a number of Gaussians, one to each diffraction order. A
typical example of the fitted momentum distributions is shown in Fig.~\ref{fitgauss}.

\begin{figure}[!t]
    \begin{centering}
    \includegraphics[width=.65\columnwidth]{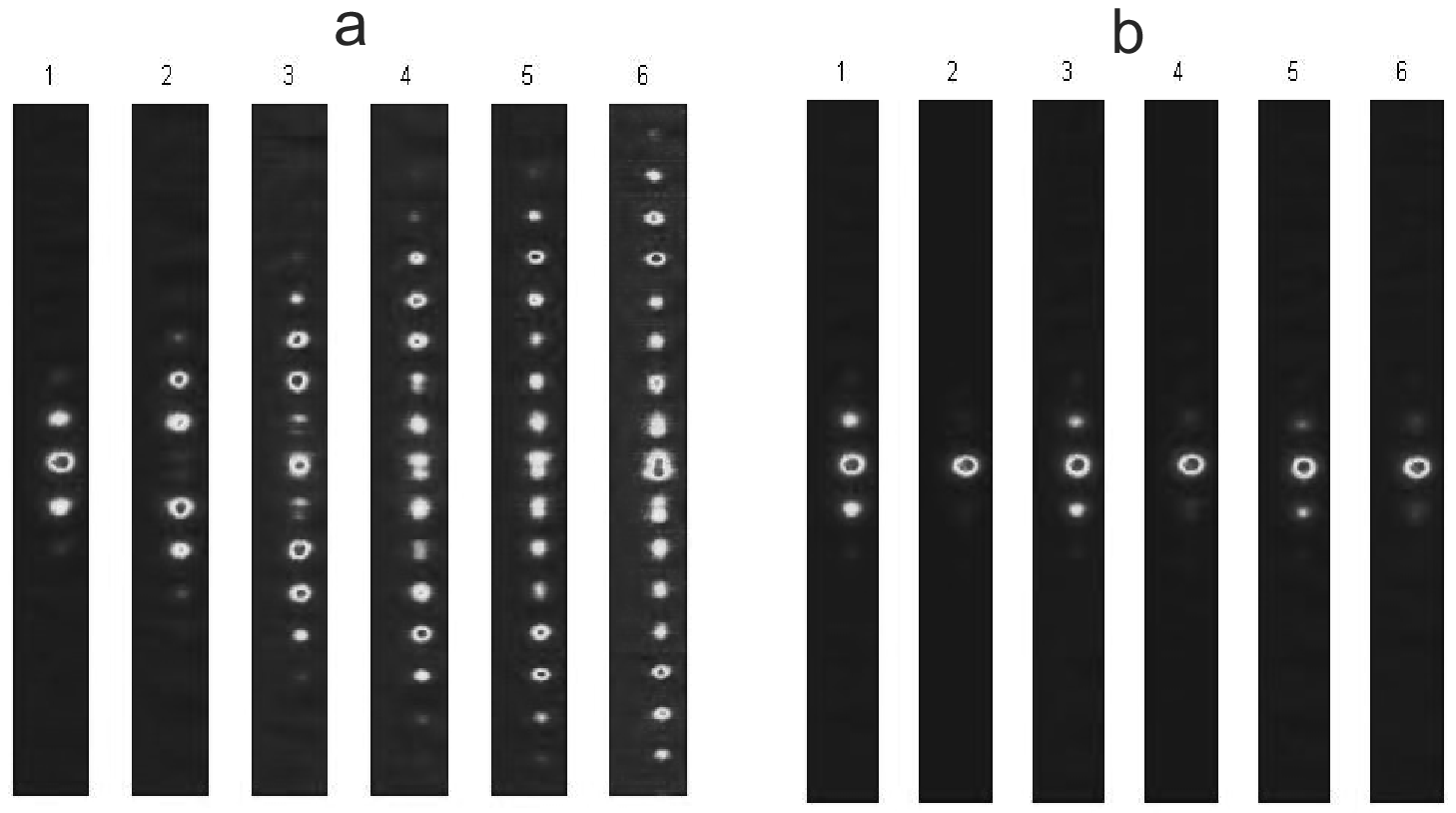}
    \includegraphics[width=.8\columnwidth]{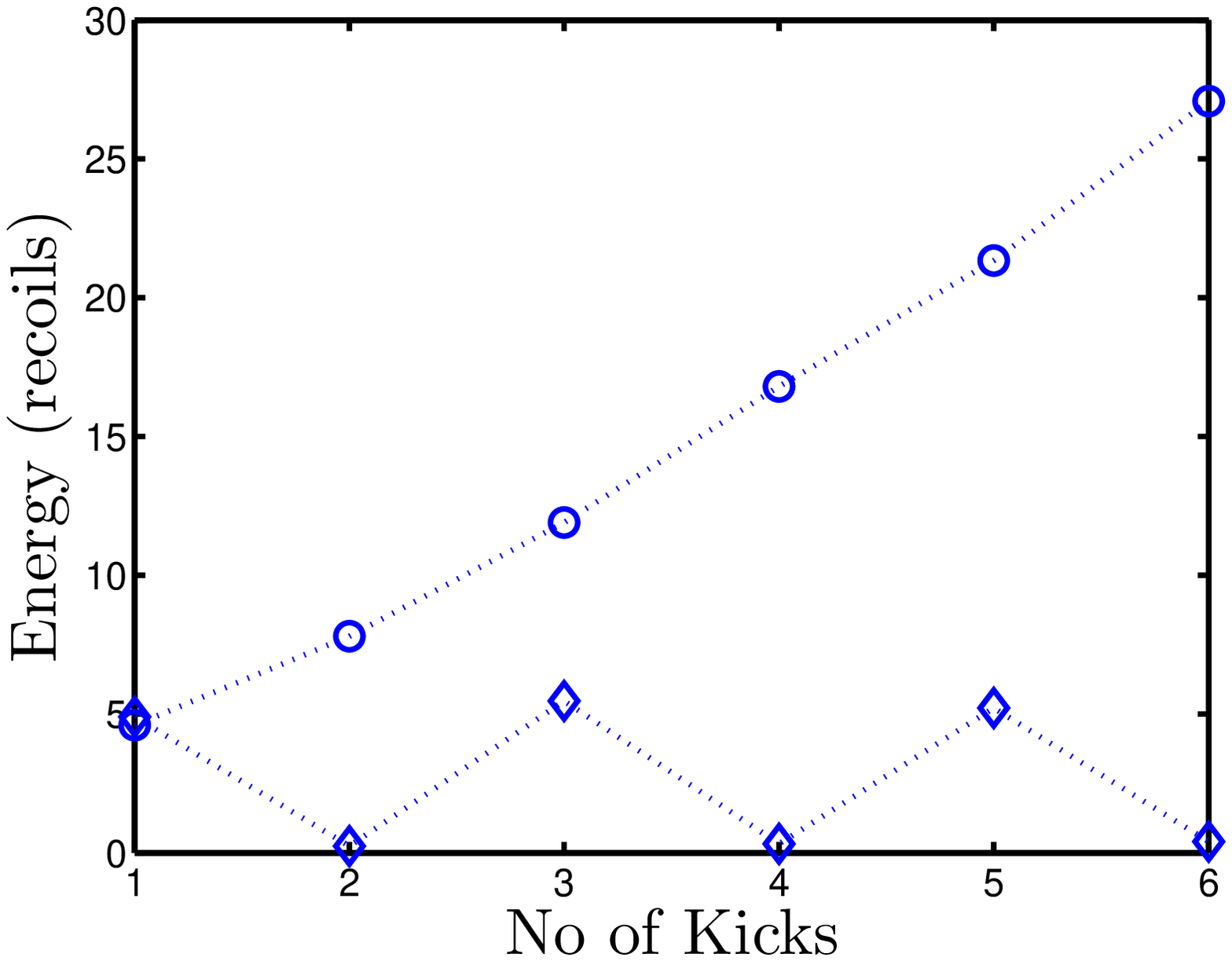}
    \caption{Experimental data: (Top) Absorption images for the quantum
resonance case (a) and the quantum anti-resonance case (b). (Bottom) Energy as a function of the number of
    kicks  ($1$ to $6$) for the quantum resonance case (circles) and anti-resonance case (diamonds). The kick strength $\phi_{d}\approx 2.5$}
    \label{resenergy}
    \end{centering}
\end{figure}
%
\subsection{Results of the primary quantum resonances}
 The kinetic energy of the system in the case of a standing wave potential $(\beta=0)$ as a function of pulse period is studied for a range of pulses or kicks. This situation can also
be thought of as equivalent to that of varying the spatial distance between the gratings in the optical Talbot effect. To start with, the energy is measured by scanning the pulse period (time
delay) for two kicks. The period between the kicks is varied from ($10-70$) $\mu$s. The second kick is used as a tool to study the time evolution of the wave function after the first kick. It is
well known that in the atomic version of the kicked rotor, at the Talbot time $T_{T}=66.3$ $\mu$s of free evolution, an identical wave function to that directly after the first kick is observed.
The introduction of another grating or kick at the Talbot time then simply doubles the effect or duration of the first kick. As a result, a quadratic growth in energy is observed
~\cite{Reichl1992,Izrailev1980}.

At a pulse period of $\sim33.1$ $\mu$s, which is equivalent to half of the Talbot time, i.e. $T_{T}/2$, the quantum phase factor between kicks alternates sign ~\cite{Oskay2000}. As a result, a
re-image of the grating's transmission function, which is the inverse of the initial one, is obtained. In effect, two kicks with half the Talbot time in between them results in the complete
negation of the effect of the first kick.
The quantum resonance and anti-resonance behavior is observed in the experiment as shown in absorption images in Fig.~\ref{resenergy} a and b (Top) with the corresponding energy of the atoms
representing these resonances (Bottom).
%
\begin{figure}[!t]
    \center
    \includegraphics[width=.8\columnwidth]{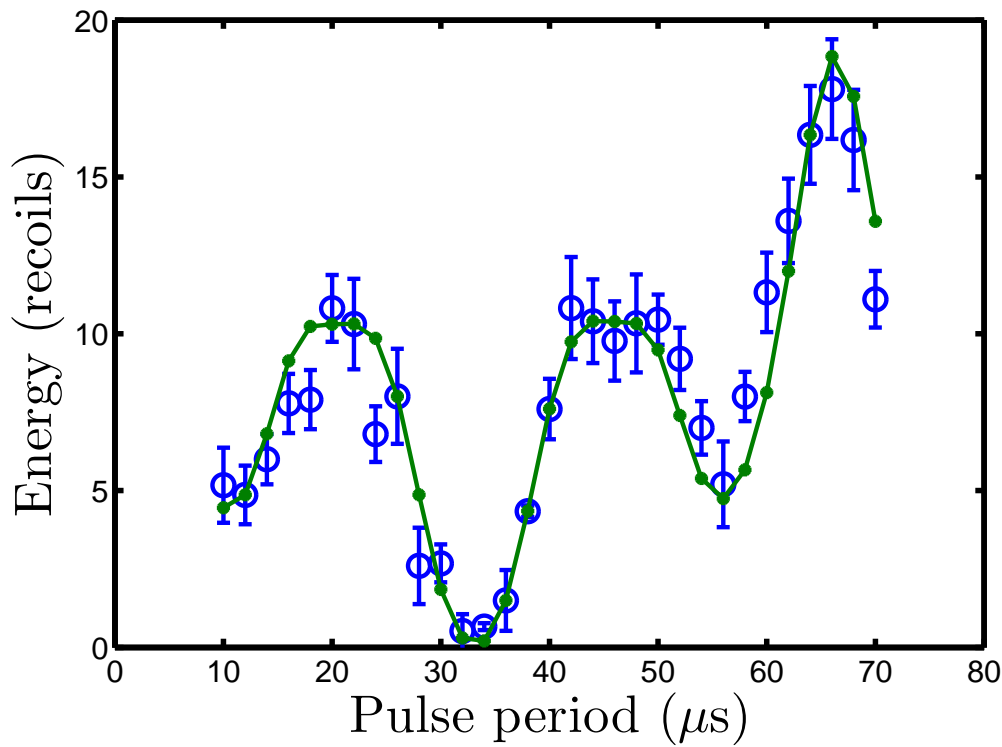}
    \caption{(Color online) The energy as a function of pulse period for two kicks. Blue circles show experimental data,
    whereas the solid green line represents the
    simulations. The parameters are $N=2$, $T=(10-70)$ $\mu$s, $\phi_{d}\approx
2.5$.}
    \label{twokicks}
\end{figure}
\subsection{Intermediate secondary resonances}
The energy dependence of the condensate diffracted by two gratings or kicks as a function of pulse period between the kicks is shown in Fig.~\ref{twokicks}. The solid curve represents the
theoretical simulations, whereas the circles are the experimental observations. The error
bars are determined by running the experiment a number of times.
The well defined peaks and valleys that are observed correspond to the integer and fractional Talbot times. The maximum in energy at $\sim66$ $\mu$s is in close resemblance with the simulation,
and is a clear indication of the Talbot time $T_{T}$ that corresponds to the quantum resonance. The minimum in energy at $\sim33$ $\mu$s is also clearly visible, matching the theoretical
predictions of half the Talbot time $T_{T}/2$.
\begin{figure}
    \center
    \includegraphics[width=.8\columnwidth]{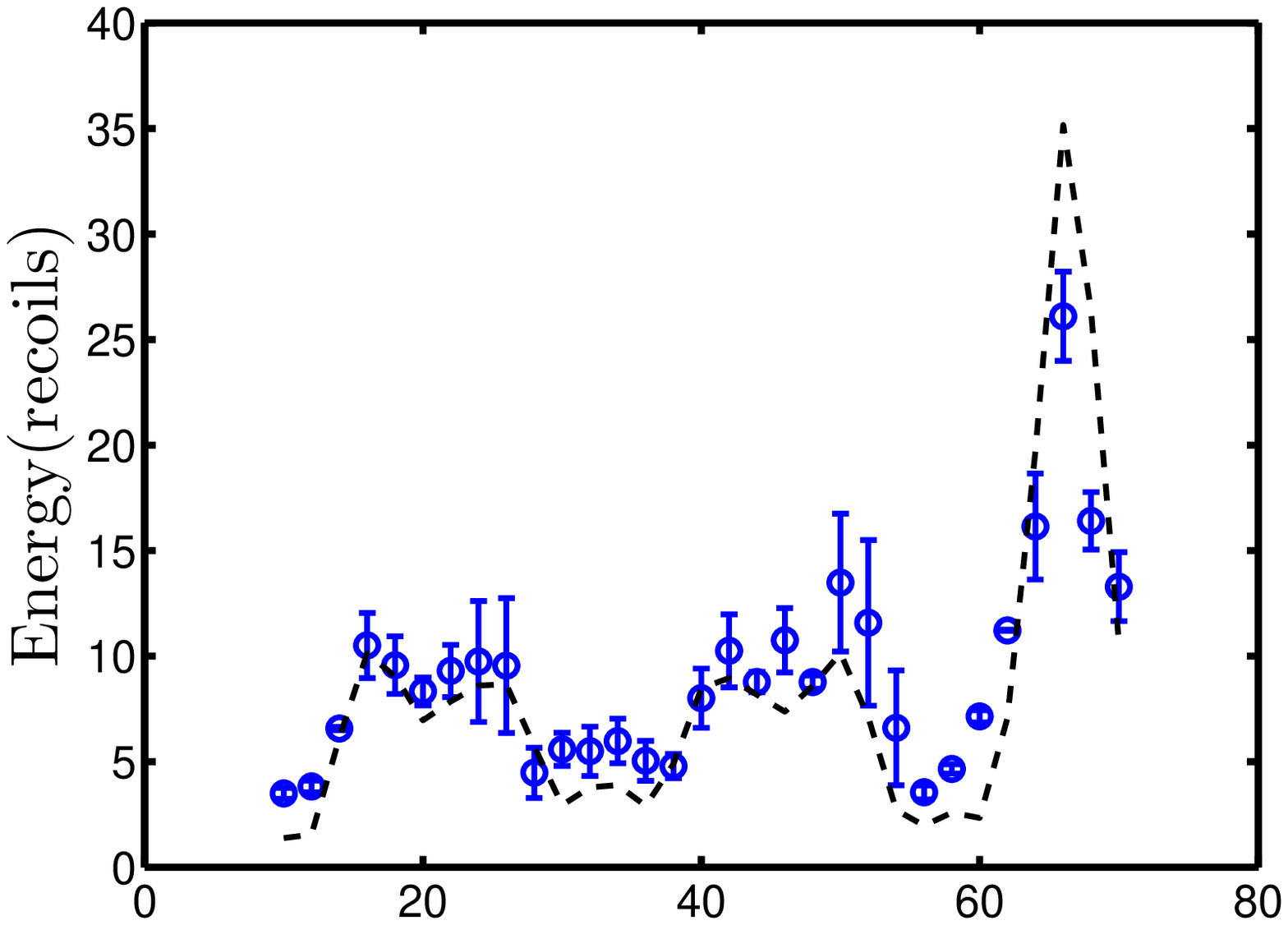}
    \includegraphics[width=.8\columnwidth]{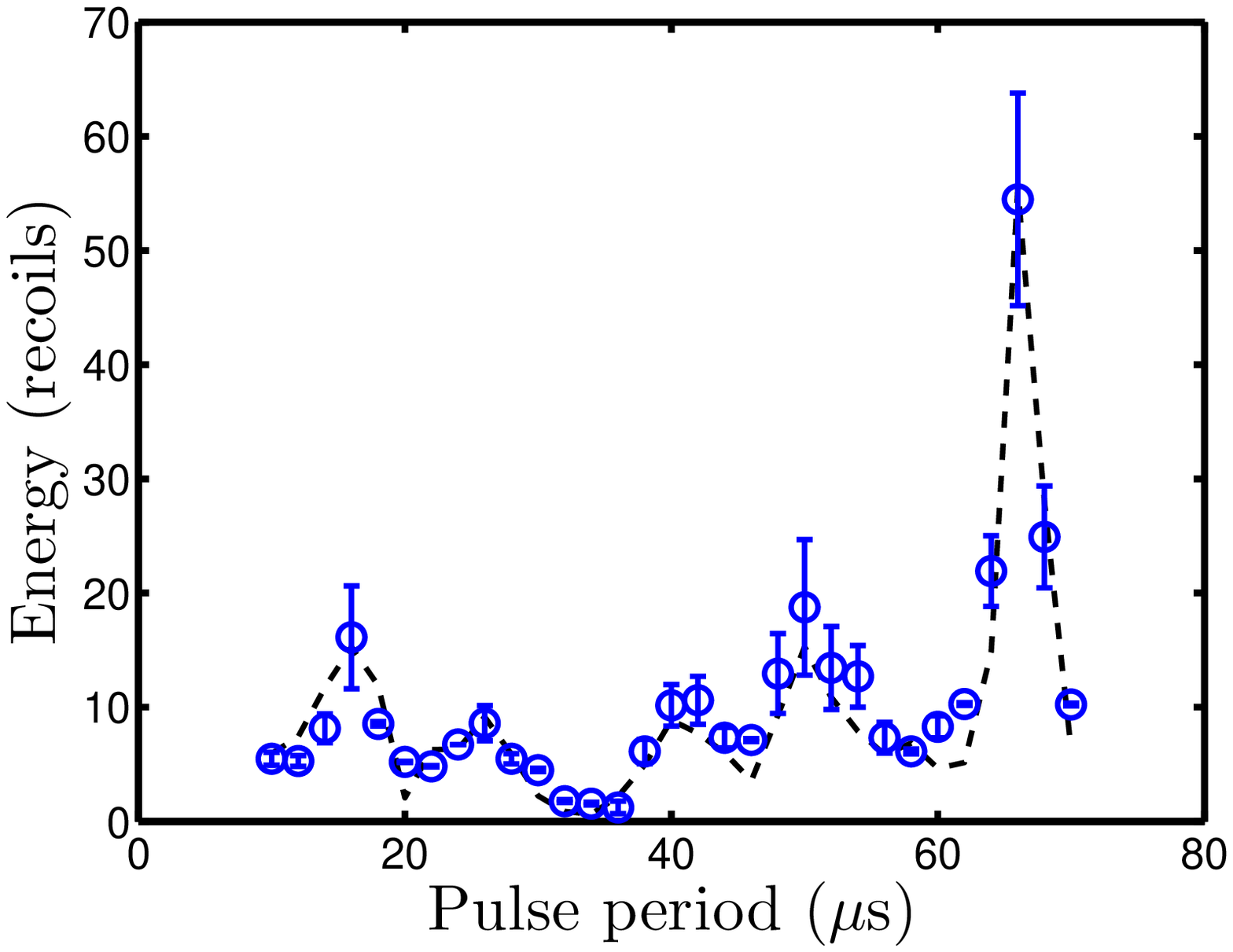}
    \caption{(Color online) The energy as a function of pulse period for three kicks (Top) and
four kicks (Bottom). The blue circles represents experimental data,
    whereas the dashed lines are results of the simulations.
    The parameters are $T=(10-70)$ $\mu$s, $\phi_{d}\approx 2.5$.}
    \label{threekicks}
\end{figure}

The energy of the system is also studied as a function of pulse period, by adding another pulse (kick) to the first two. The addition of this pulse increases the energy of the system by inducing
an extra phase shift. The global minimum in this case is then replaced by a local maximum as the complete cancelation of the first two kicks is destroyed by the third one. This is true for an
odd number of kicks. The final energy of the system for three kicks as a function of pulse period is shown in Fig.~\ref{threekicks} (Top). Most of the structure in Fig.~\ref{threekicks} is less
visible; however, the maximum in energy at about $T_{T}$ is still obvious as the gratings for this amount of free evolution do not induce any additional phase shift.

\begin{figure}
    \center
    \includegraphics[width=.8\columnwidth]{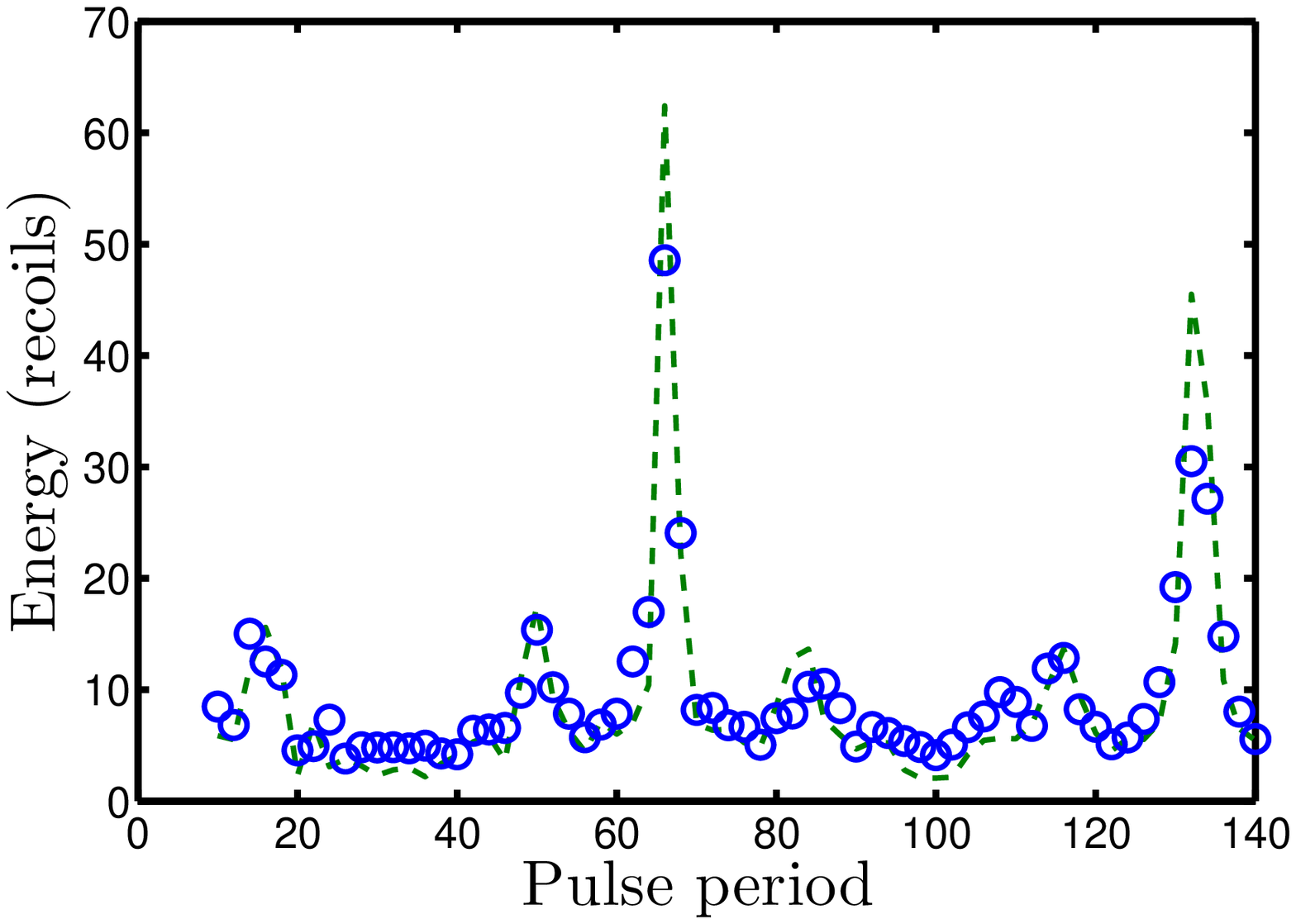}
    \caption{(Color online) The energy in recoils as a function of pulse period. Experimental data (blue circles),
    whereas the dashed lines are results of the theoretical simulations. The
parameters are $N=5$, $\phi_{d}\approx 3$ and $T=(10-140)$ $\mu$s.}
    \label{fivekicks}
\end{figure}

The evolution of the system in the presence of large number of kicks shows interesting results in the simulations. By measuring the energy, the evolution is examined in the presence of four and
five kicks. It is well known that with an increase in the number of kicks, the width of the quantum resonances when plotted against the pulse period becomes narrow ~\cite{Currivan2009,Ryu2006}.
In Fig.~\ref{threekicks} (bottom), the energy as a function of pulse period is plotted for four kicks. In this case, at a pulse period of $\sim T_{T}/2$, the minimum in energy (close to zero) is
observed again. As for an even number of kicks, the effect of the previous kick is canceled again for the corresponding time delay of $T_{T}/2$, and almost no diffraction occurs.
The maximum in energy at $T_{T}$ is again visible, since for this time delay the energy of the system increases with the addition of further kicks. It should be noted that the resonance peak
observed at $T_{T}$ is narrow compared to the peak for a small number of kicks. This is in accordance with theoretical simulations, where a narrowing of the quantum resonance peaks for a large
number of kicks is observed.

A similar behavior is observed when an extra pulse is added. In Fig.~\ref{fivekicks}, the energy is examined by a complete scan of the pulse period up to the second primary resonance for five
kicks. The minimum in energy at $T_{T}/2$
, which was obvious in the case of four gratings, however, is replaced by a local maximum as shown. This is because of the extra phase shift induced by the fifth grating, due to which the energy
increases by some amount for the corresponding time delay. It is shown that much narrower resonance peaks are observed in this case around the quantum resonances of $T_{T}$ and 2$T_{T}$
($\sim132$ $\mu$s). The maximum in the experiment in this case, however, is not in agreement with the simulation. We believe this discrepancy is caused by the limited signal to noise ratio.
Because of the limited resolution of the experiment, the observation of the probability of the atoms at high momenta and detecting a small amount of diffraction orders at high momenta is
difficult, which results in the reduction of the energy.
Because of finite width of the initial momentum distribution the energy of the momentum distribution tends to be extremely sensitive to the population at high momenta ~\cite{Oskay2000}. Overall,
the experimental results are in good agreement with the theoretical predictions and simulations.
\begin{figure}[!t]
\begin{center}$
\begin{array}{cc}
\includegraphics[width=.45\columnwidth]{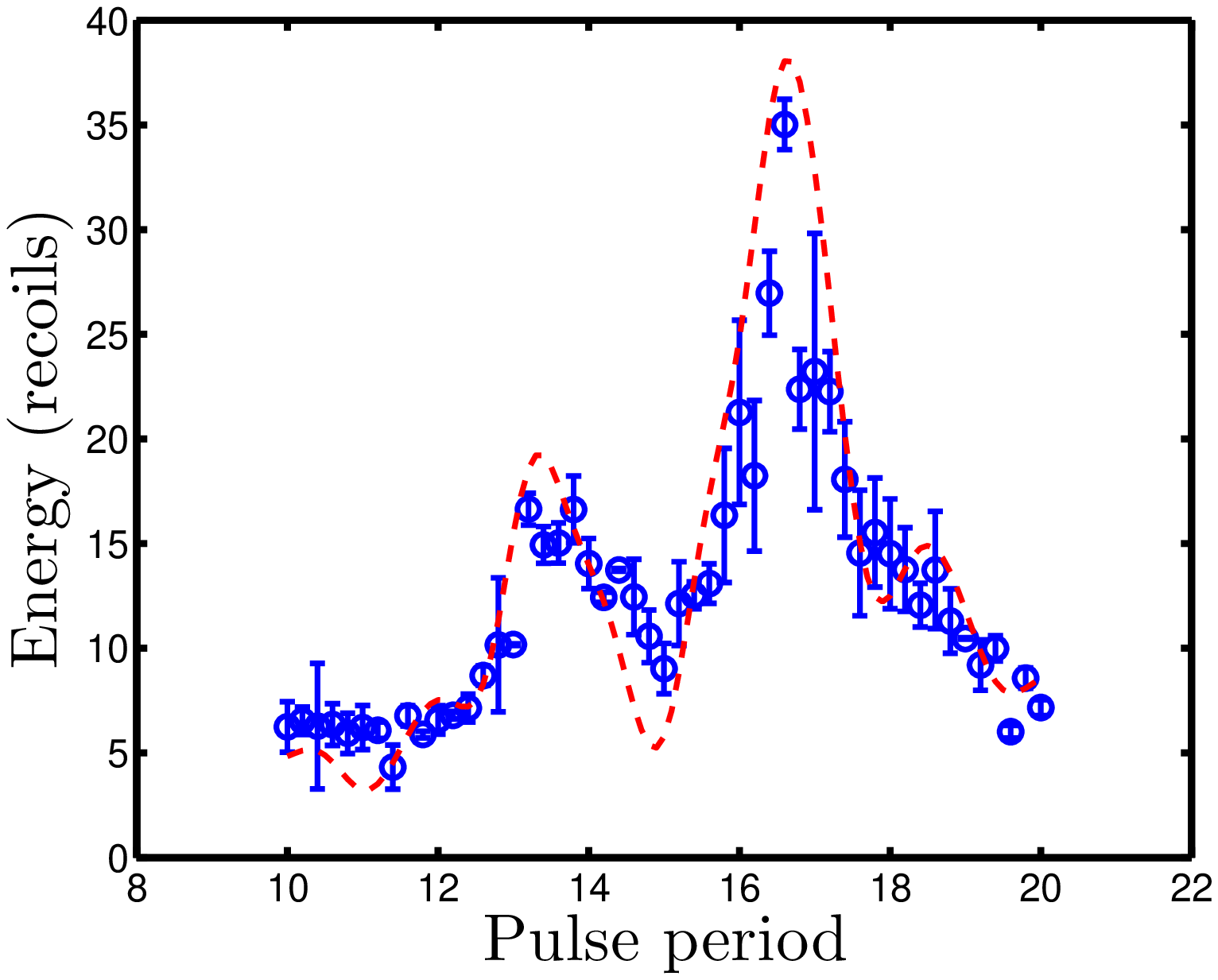} &
\includegraphics[width=.53\columnwidth]{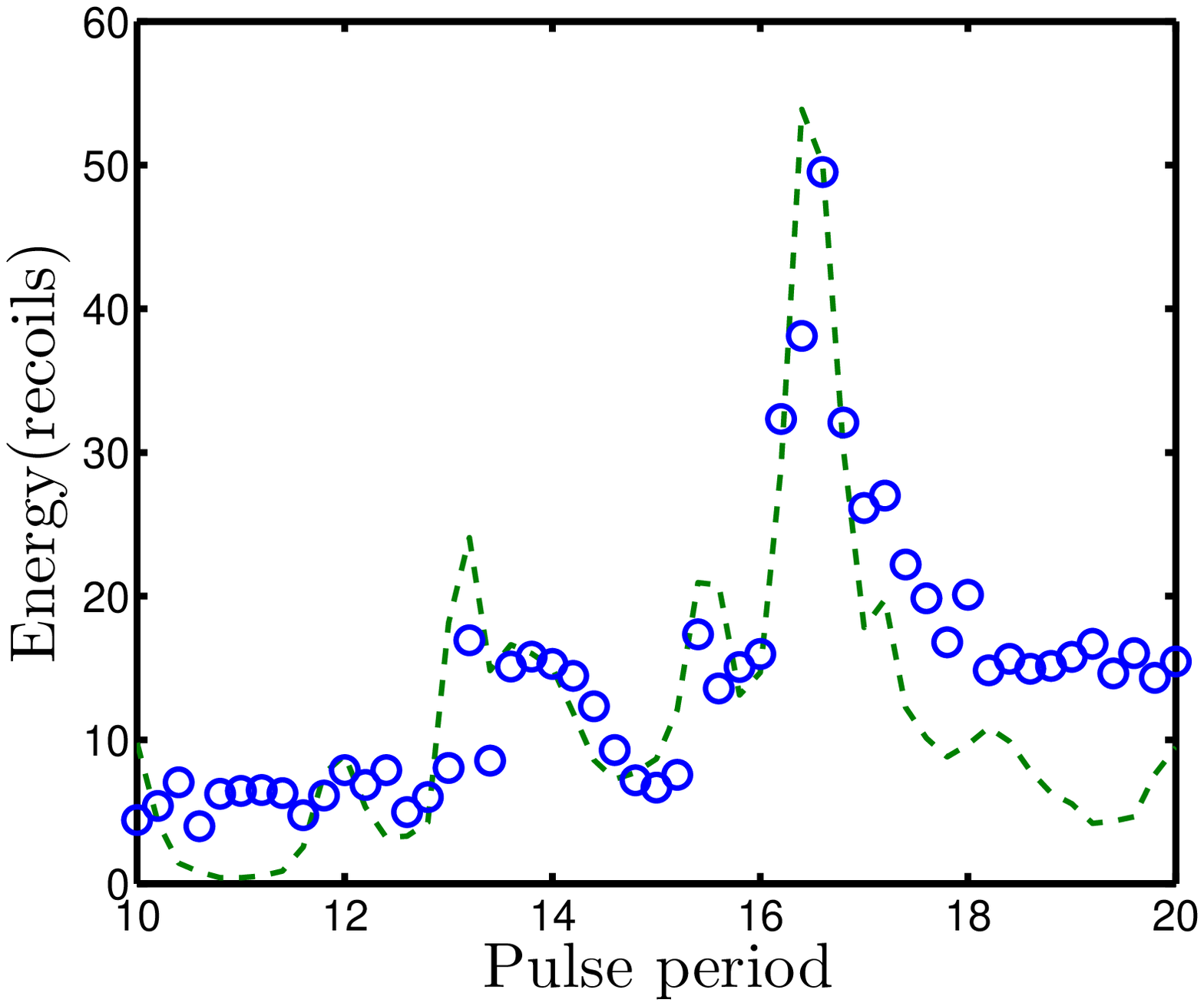}
\end{array}$
\end{center}
\caption{(Color online) The energy in recoils as a function of pulse period or time delay between the gratings is shown for five (left) and ten (right) kicks with a kick strength of
$\phi_{d}\approx 3$. Blue circles show experimental data, whereas dashed lines represents the simulations.}
 \label{frac}
\end{figure}
%
%
Higher order resonances are expected to occur in these systems at fractional multiples of the Talbot time $T_{T}$, i.e., at
\begin{equation}
\ T_{frac}=\left(\frac{l}{m}\right) T_{T},
\end{equation}
where $l$ and $m$ are integers forming a rational number. The quantum revivals, or resonances, at these fractional times are the results of superposition of $q$ copies of the initial wave
packet, which are separated by $2\pi/q$ ~\cite{Berry2001}. Expecting resonances at these fractional times, the energy of the atoms is measured and plotted as a function of pulse period in
between the gratings. In Fig.~\ref{frac} the energy is shown for a full scan of the pulse period from $(10-20)$ $\mu$s for $5$ and $10$ kicks. The maximum in energy at $\sim 16.7$ $\mu$s is the
evidence of a fractional resonance at $(1/4)T_{T}$. Similarly, the small peak at  $\sim 13.2$ $\mu$s shows another resonance which corresponds to a fractional Talbot time of $(1/5)T_{T}$.

In Fig.~\ref{frac} (right), the fractional resonances become sharp as a result of large number of kicks, ten kicks in this case. For multiple gratings, the energy of the atoms is more sensitive
to each induced phase shift and it is extremely difficult to analyze accurately in the experiment. The experimental data follow the general trend of those from the simulations.

\section{Initial momentum dependence on quantum resonances}
In this section, measurements of the effect of the initial momentum on the primary quantum resonances in the delta-kicked rotor are discussed. The study of quantum transport in such systems has
been of great interest in recent years in experiments with cold atoms ~\cite{Sadgrove2007,Dana2008}, but its dependence on the initial velocity of the atoms still needs further investigation.

\subsection{Experimental setup}
\begin{figure}
\center
\includegraphics[width=.8\columnwidth]{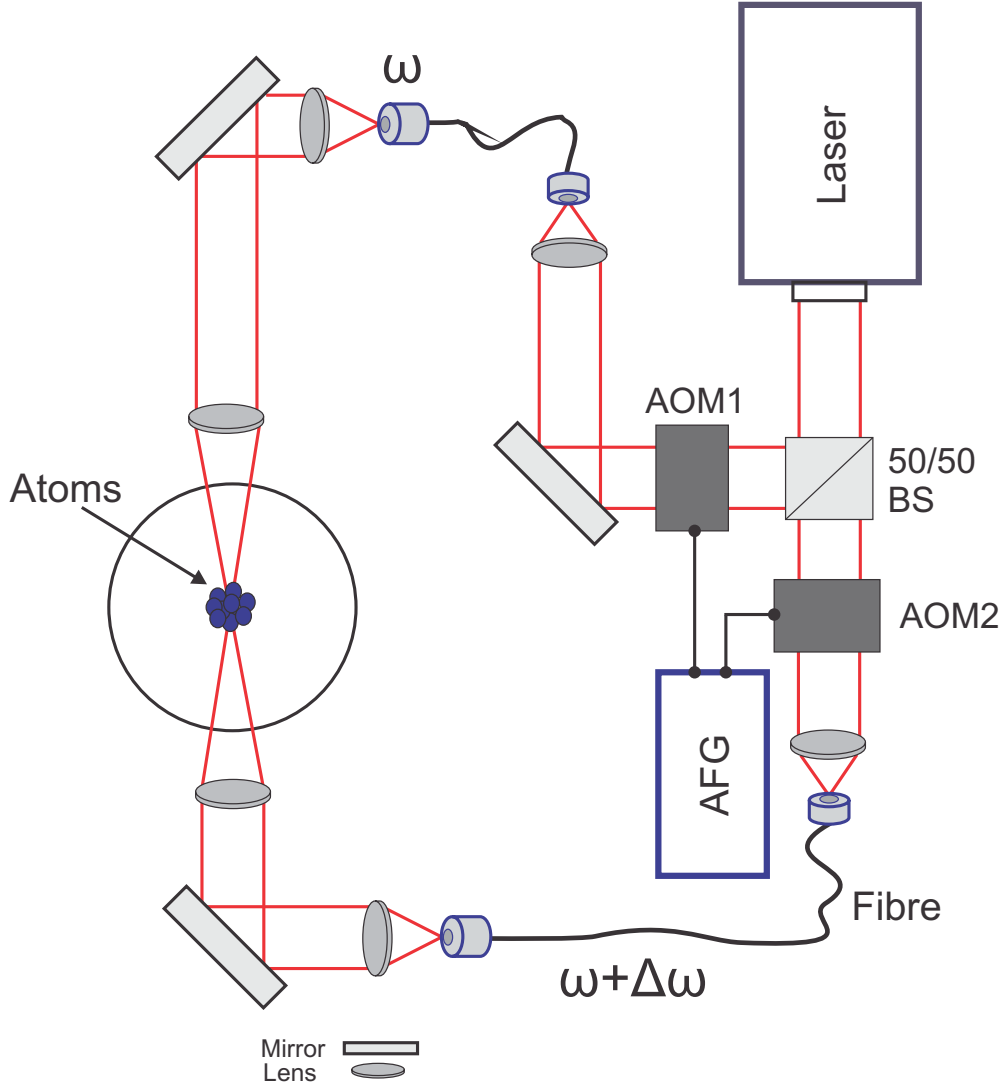}
\caption{A sketch of the laser setup for a moving standing wave: The AFG (Arbitrary Function Generator) is used to drive the AOMs, enabling us to change the frequency of one of the laser beam by
a small amount.} \label{kicksetup}
\end{figure}

In order to take into account the initial momentum dependence on the quantum resonant effects, a modified kick laser setup is used, shown in Fig.~\ref{kicksetup}. The laser beam is split into
two by a beam splitter, and the two beams are then passed through separate AOMs. The AOMs used for fast switching are driven in this case by an Arbitrary Function Generator (Tektronix AFG-3252),
amplified by
home-built RF amplifiers. 
The AOMs are switched on simultaneously, with a tunable frequency difference $\delta \omega$. The AFG generates two $80$ MHz (plus a desired offset frequency), $3.5$ V$_{pp}$ sine waves that
shift the frequency of the laser beams. The frequency difference gives us an effective initial momentum $p_{i}$ for the atoms, given by $p_{i}/p_{rec} = \delta \omega/(4\omega_{rec})$.
 In the experiment, using the definition of $E_{rec}=\hbar^{2}k^{2}/2m$, for
$\lambda=780$ nm one recoil frequency is given by $f_{rec}=3.77$ kHz.
 From the definition above and in
~\cite{Saunders2007}, $3.77$ kHz corresponds to a value of the quasimomentum $\beta=.25$ recoils, and $30.16$ kHz corresponds to $\beta=2$ recoils. An equation editor ArbExpress was used to
generate a pulse sequence for different pulse periods. The frequency of one beam was set at $80$ MHz and the second beam was varied by $80$ MHz+$3.77$ kHz to $80$ MHz+$30.16$ kHz in eight steps.
By introducing the small frequency difference, a moving wave rather than a standing wave was obtained. The frequency difference corresponds to quasimomentum values of zero to two photon recoils.
It should be noted that what is important is the relative phase of the sine waves driving the AOMs in the subsequent pulses, caused by the small frequency difference, not the actual frequency
difference.

An ensemble of cold atoms with a momentum spread much less than a single photon
recoil is used. These cold atoms are then kicked at kick periods
corresponding to quantum resonances. The
analytical expression for the amplitudes of the momentum states with momentum
$2j\hbar k$ after $n$ kicks has been derived in ~\cite{Wimberger2003}
and ~\cite{Saunders2007}, for the kick periods
that are half integer times the Talbot time $T_{T}$,
\begin{equation}\label{ch4-eqn-cjvalues}
\ c_{j}=J_{j}\left(\phi_{d}\frac{\sin(n\Upsilon)}{\sin\Upsilon}\right) i^{j} e^{-\iota j(n+1)\Upsilon}e^{-i n \pi \beta^2 l},
\end{equation}
where $\beta$ is the quasimomentum in units of $2\hbar k_{L}$, and $l$ is an
integer, given by $l=2 T/T_{T}$. The parameter $\beta$ can be used to assign
different initial momenta to the atoms
in the experiment. From equation 16 and 19, the second order moment of the
momentum distribution can be found as
\begin{equation}
\left\langle p^{2}\right\rangle= (2\hbar k_{L})^2 \sum\limits_{j=-\infty}^\infty
|c_{j}|^2 (j+\beta)^2.
\end{equation}
The values obtained from Eqn. (\ref{ch4-eqn-cjvalues}) are found to be in
excellent agreement with those obtained from the simulation, and with the
experimental data.
\begin{figure}[!t]
       \center
       \includegraphics[width=.8\columnwidth]{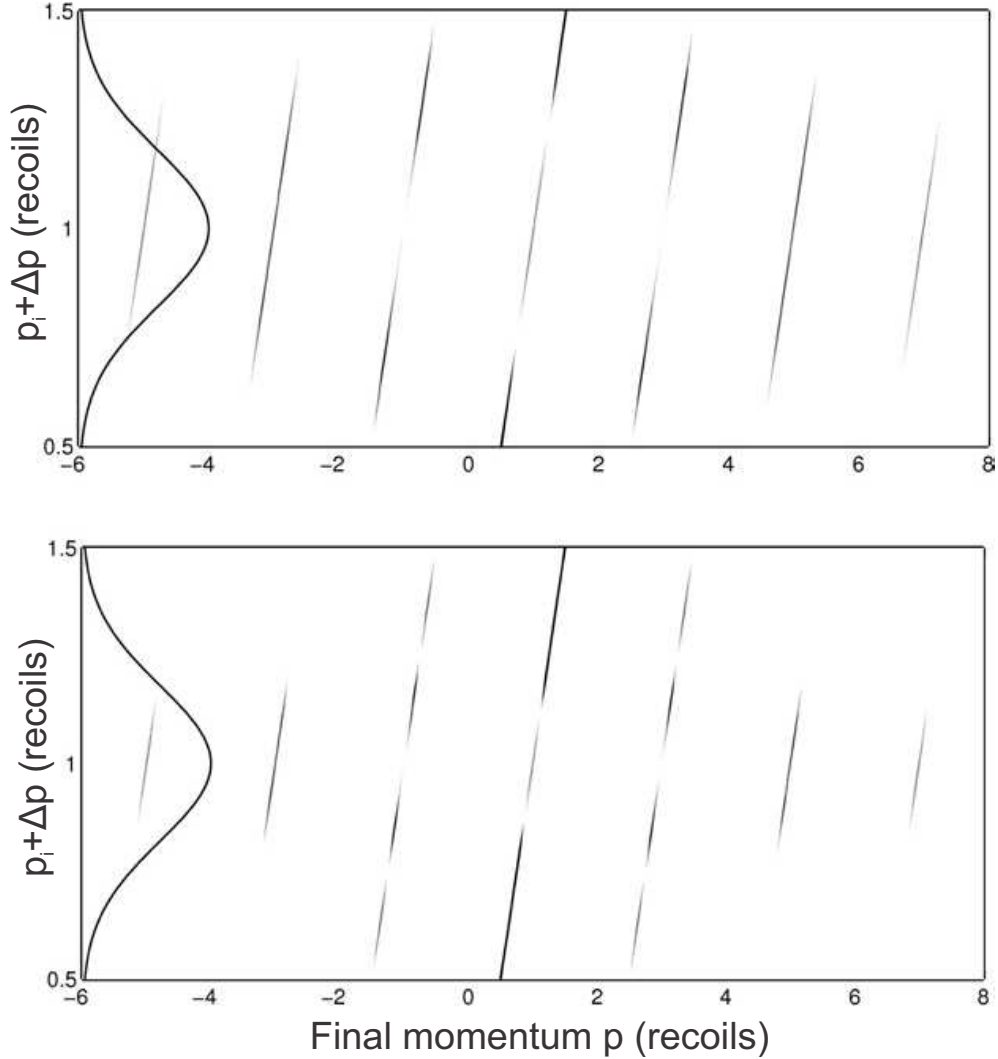}
       \caption{The probability of finding atoms at the final momenta
(horizontal axis) for a range of initial momenta (vertical axis) in the
       simulation. The probability for kick period
       of $T=T_{T}/2$
       (Top) and $T=T_{T}$ (Bottom) is shown for four kicks, with average
initial momentum $\left<p_{i}\right>=p_{rec}$ and $\phi_{d}\approx 1$.
       The original BEC distribution is also shown
       (as solid curve).}
       \label{deltap}
\end{figure}

\subsection{Results and discussions}
\begin{figure}[!t]
       \center
       \includegraphics[width=.6\columnwidth]{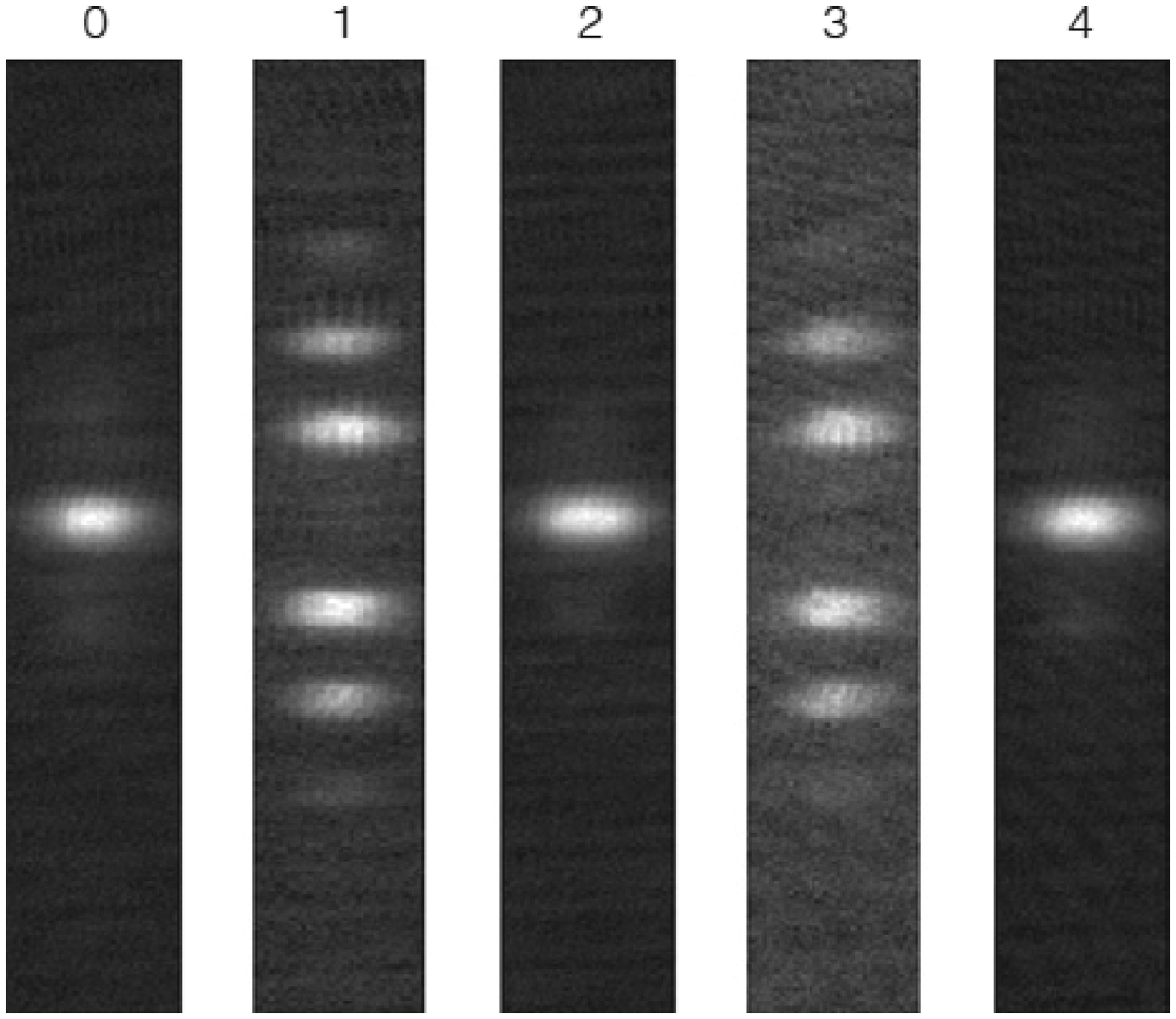}
       \caption{ Absorption images of the atoms after two kicks, with a kick
period of $T=T_{T}/2$ for different initial momenta from $0$ to $4$ recoils.}
       \label{ch4abs}
\end{figure}
 In Fig.~\ref{ch4abs}, absorption images of the atoms are shown for the
antiresonance $T = T_{T}/2$ case after two kicks for different initial momenta.
 The atoms are
allowed to expand for about $5$ ms before they are imaged by the camera, using the time-of-flight method. The different initial momentum is imparted to the atoms by setting a small frequency
difference between the two laser beams. At $p=0$ the effect of the first kick is exactly canceled by the second for the kick period corresponding to half the Talbot time, and we see almost no
diffraction, as expected from the Talbot effect. At the initial momentum of $p_{i} = p_{rec}$, we see strong diffraction, as $T_{T}/2$ corresponds to a resonance for these atoms. A similar
effect is observed for atoms with an initial momentum of $p_{i} = 3 p_{rec}$. For atoms with an initial momentum of $p_{i} = 2 p_{rec}$ and $p_{i} = 4 p_{rec}$ there is again no diffraction. The
reason for the absence of diffraction in this case is due to the fact that $T_{T}/2$ corresponds to an antiresonance for these atoms. For the antiresonance case here, the free evolution part of
the Floquet operator evolve according to
\begin{equation}
\ U_{free}= e^{-i (2N+1)\frac{\hbar^{2}k^{2}}{2m}T_{T}/2},
\end{equation}
and depends on the initial momentum. For $T_{T}/2$ the free evolution operator depends on $\hbar k$, resulting in resonances and antiresonaces for specific values as described. The laser field
induces coupling only between momentum eigenstates differing in momenta by integer multiples of $\hbar k$ ~\cite{Saunders2007}.
\begin{figure}[!t]
       \center
       \includegraphics[width=1\columnwidth]{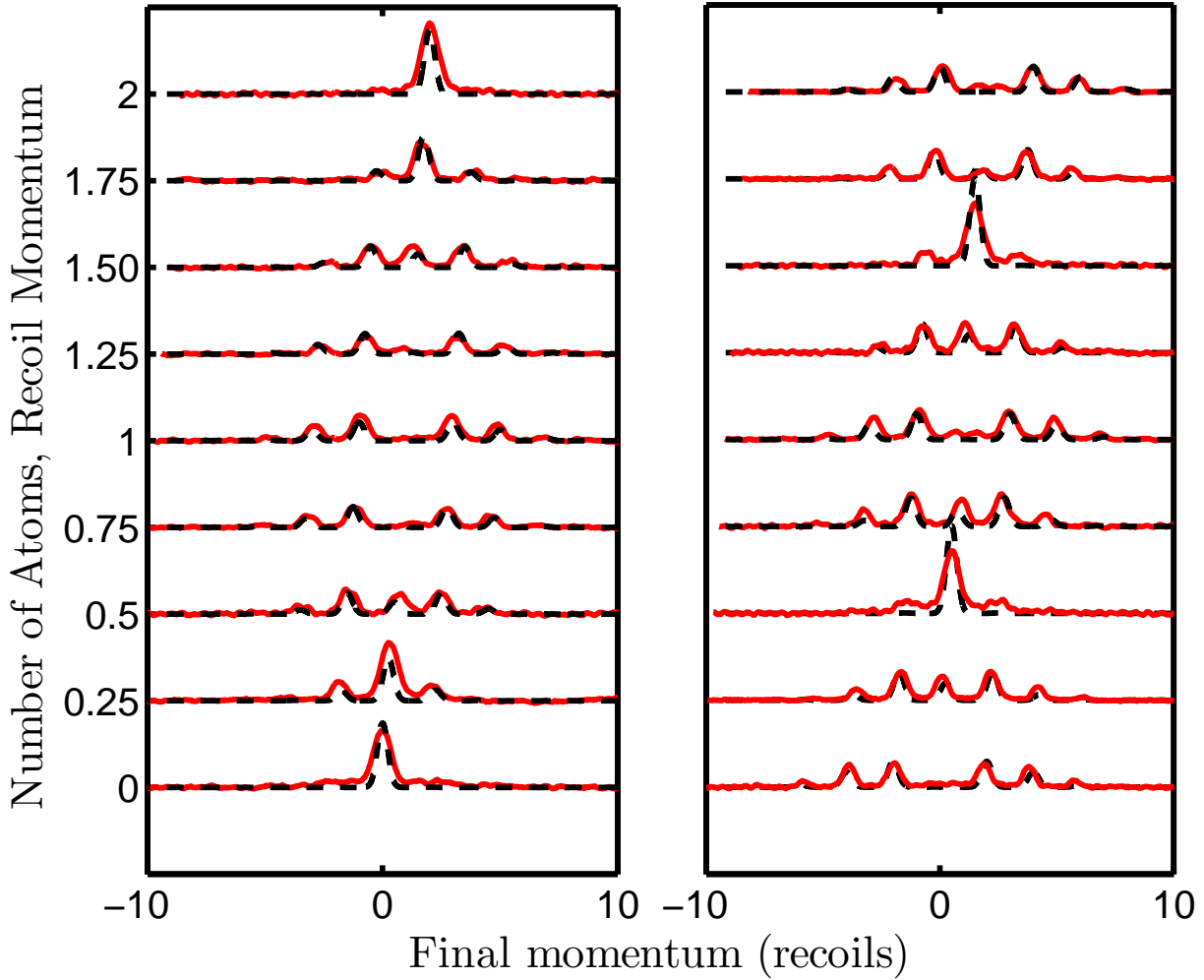}
       \caption{(Color online) The number of atoms as a function of the final momentum, for a
range of initial momenta $0$ to $2\hbar k$. On the left we show
       the experimental results for two kicks with a
       period of
       $T_{T}/2$. On the right the results are shown for two kicks with a period
of $T_{T}$. The solid red curves represents the experimental data,
       whereas the simulations are shown as dashed lines, for a kick strength of $\phi_{d}\approx 1$.}
       \label{fig2all}
\end{figure}
%

The dynamics of the system are examined for a range of initial momentum at pulse periods of $T_{T}/2$ and $T_{T}$ between the kicks. In Fig.~\ref{fig2all} on the left, the momentum distribution
of the atoms after the kick sequence is examined for the initial momentum $p_{i}$ in 8 steps, from 0 to 2 $p_{rec}$. The resulting 1D momentum distributions, averaged over the three repeats of
the experiment, are shown as a function of the final momentum for two kicks. It can be seen that at zero velocity, there is indeed an anti-resonance and very little diffraction occurs.
For small increments in the initial momentum, small peaks start to appear at higher diffraction orders. At $p_{i}=0.5 p_{rec}$, the first and second order diffraction peaks are observed. At an
initial momentum of $p_{i}=1p_{rec}$, the evolution turns into a resonance, with significant diffraction. The probability of atoms at higher momenta starts to decrease for $p_{i}>1p_{rec}$ and
at $p_i = 2p_{rec}$ the system returns back to an anti-resonance, with very little diffraction. Also shown in the figure are the results of the simulation. When experiments are performed with a
broad initial momentum distribution, an average over many different $v_{i}$ yields an observed resonance at $T = T_{T} /2$, as has been shown in many publications, see e.g. ~\cite{Oskay2000}.

On the right in Fig.~\ref{fig2all}, the resulting 1D momentum distributions in the experiment and the simulations are shown for a period between the kicks of $T = T_{T}$. The initial momentum
$p_{i}$ is again varied from $0$ to $2$ $p_{rec}$ in 8 steps. At an initial momentum of $p_{i} = 0$, there is a resonance and strong diffraction occurs as expected. The evolution of the system,
however, turns into an antiresonance at $p_{i}=0.5 p_{rec}$ with small diffraction, and back to a resonance at $p_{i}=1 p_{rec}$. The probability of atoms occupying the higher momentum states
start to decrease again, and at $p_{i} = 1.5 p_{rec}$ an antiresonance is observed. The evolution finally turns  into a resonance again at $p_{i} = 2 p_{rec}$, where a strong diffraction is
seen. Again, averaging over a range of initial velocities will show the same resonance at $T = T_{T}$ , even though the dependence on the initial velocity cycles at twice the rate. At $T = 3
T_{T}/2$, the cycle of the amount of diffraction varying with the initial velocity is at three times the rate we see for $T= T_{T}/2$, and so on. The simulated curves show good agreement in
terms of the relative heights of all the diffraction peaks.
\begin{figure}[!t]
       \center
       \includegraphics[width=.8\columnwidth]{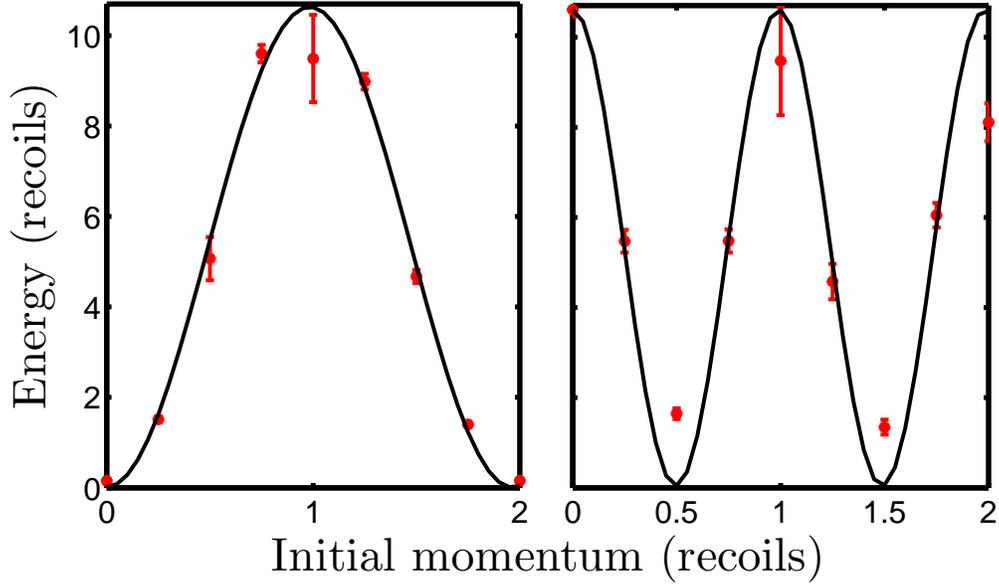}
       \caption{The final energy in recoils as a function of initial momentum
from $0$ to $2\hbar k$ for $T=T_{T}/2$ (left) and $T=T_{T}$ (right)
       is shown for the experiment with uncertainties
       (as markers) and the simulation (as solid line) for two kicks, with
$\phi_{d}\approx 1$.}
       \label{fig3aplusb}
\end{figure}
\subsection{Effect of the initial momentum on the kinetic energy}
As discussed earlier, the AOKR dynamics can be examined by the observation of the energy or the variance of the momentum distribution in units of the recoil energy $E_{rec}=\hbar \omega_{R}$. In
this section, we study how the energy of the system is influenced by taking into account the initial momentum of the atoms. The energy is scanned as a function of the initial momentum $p_{i}$
for different pulse periods for a number of kicks. In Fig.~\ref{fig3aplusb}, the variation of the energy on the initial momentum for $T = T_{T}/2$ (left) and $T = T_{T}$ (right)
%
\begin{figure} [!t]
       \center
       \includegraphics[width=.8\columnwidth]{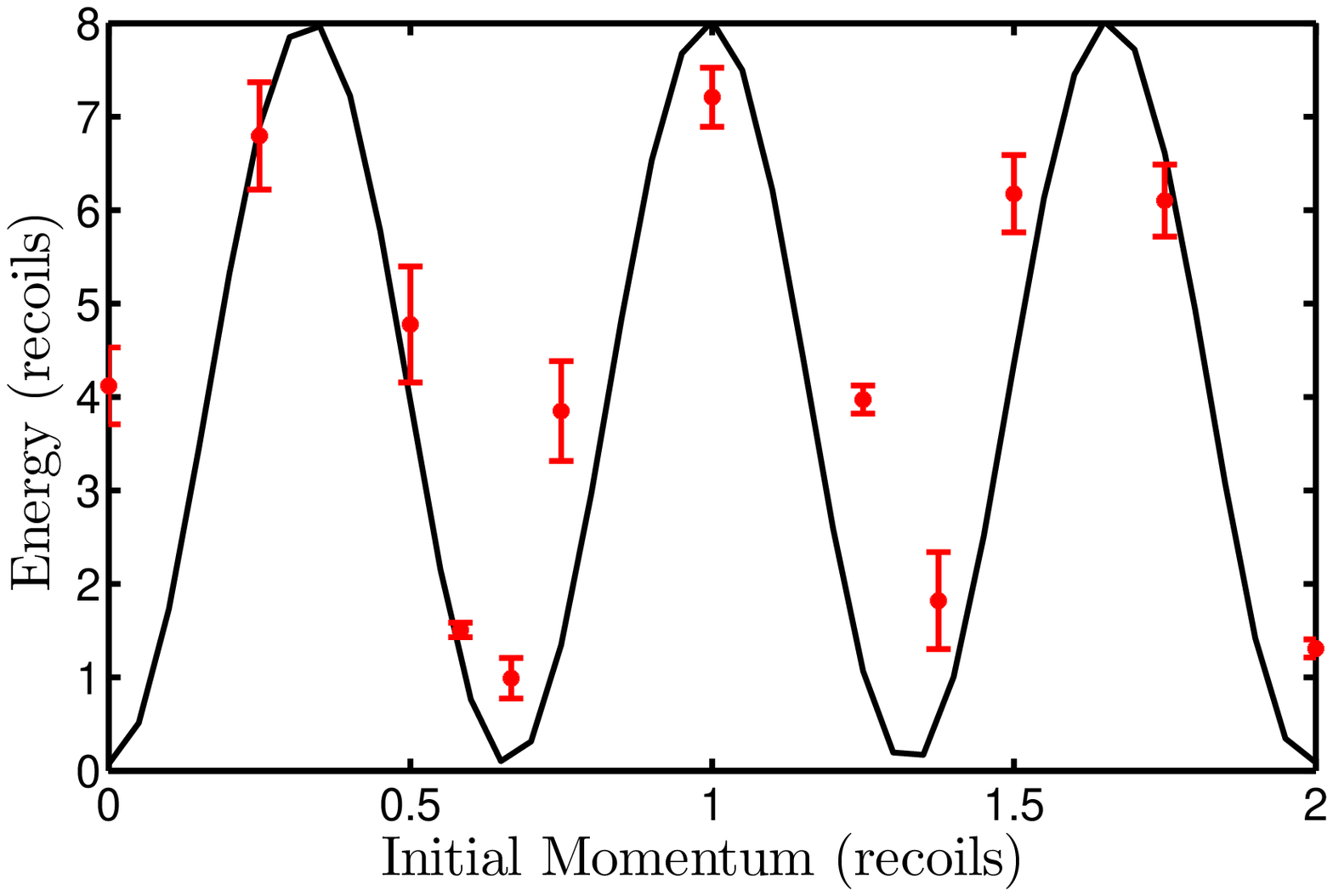}
       \caption{The final energy in recoils as a function of initial momentum
from $0$ to $2\hbar k$ for $T=3 T_{T}/2$ is shown for the experiment with
uncertainties
       (as markers) and the simulation (as solid line) for two kicks. The kick
strength $\phi_{d}\approx 0.9$.}
       \label{figthreec}
\end{figure}
in both the experiment and simulation for two kicks is shown. For $T=T_{T}/2$ we see a strong peak in the energy at about $p_{i}=1 p_{rec}$. 
Similarly, for the
resonant case $T=T_{T}$ oscillation in the energy is observed. These results are consistent with the results presented in Fig.~\ref{fig2all}. For $T = 3T_{T}/2$, the energy varies at three times
the rate for $T = T_{T}/2$, as shown in Fig.~\ref{figthreec}. For the data points, the averaged profile is fitted to a number of Gaussians, one for each diffraction order, to obtain the energies
after the kick sequence in each run. 
The standard deviation is then taken to estimate uncertainties in the results. A direct numerical variance of the velocity distribution was found to give similar results. Because of the
sensitivity to small noise peaks at large momenta, these results were less consistent, and were therefore not used.

\begin{figure}[!t]
       \center
       \includegraphics[width=.9\columnwidth]{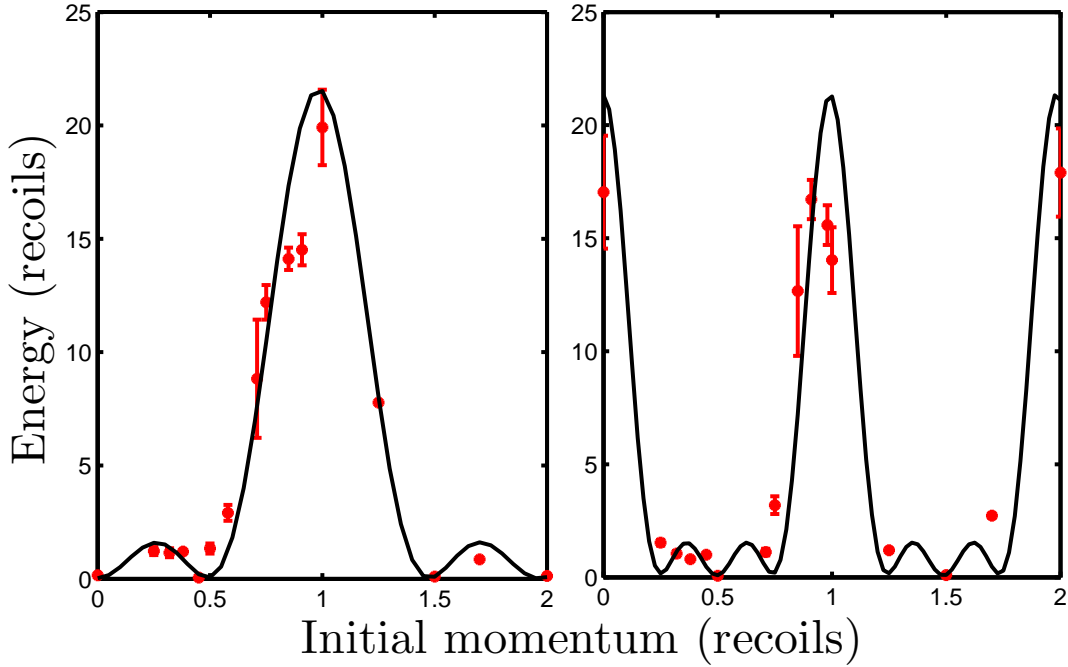}
       \caption{(Color online) The final energy in recoils as a function of initial momentum
from $0$ to $2\hbar k$ for $T=T_{T}/2$ (left) and $T=T_{T}$ (right) is shown for
the experiment with uncertainties
       (as red markers) and the simulation (as solid line) for four kicks, with $\phi_{d}\approx 1$.}
       \label{fig3dpluse}
\end{figure}

Next, the energy is examined as a function of the initial velocity for four kicks. As found, for a large number of kicks, the widths of quantum resonances when plotted as a function of kick
period get smaller as the number of kicks get larger. Fig.~\ref{fig3dpluse} shows the energy as a function of the initial velocity for four kicks, for both the anti-resonance $T=T_{T}/2$ (left)
and the resonance case $T=T_{T}$ (right). On the left, for $T = T_{T}/2$ there is an anti-resonance at zero momentum, yielding very small energies. There is a small maximum at an initial
momentum of $1/4$ recoils, decreasing again close to zero at $1/2$ recoils. At an initial momentum of one recoil, there is a stronger maximum in the energy, decaying again to small energies at
$3/2$ recoils before another smaller maximum. At initial momenta of two recoils, very small diffraction occurs and the energy returns to a small value close to zero. It should be noted that all
these features, prominent in calculations, can be reproduced in the experiment.
For the kick period corresponding to $T = T_{T}$, a larger number of oscillations in energy in the simulation is observed. The maximum in these curves is observed to be much narrower than those
for a small number of kicks. The maxima in the experiment is in agreement with that in the simulation, but the small-period oscillations in the simulation were not resolved in the experiment due
to limited resolution for these settings.
\begin{figure}[!t]
       \center
       \includegraphics[width=.9\columnwidth]{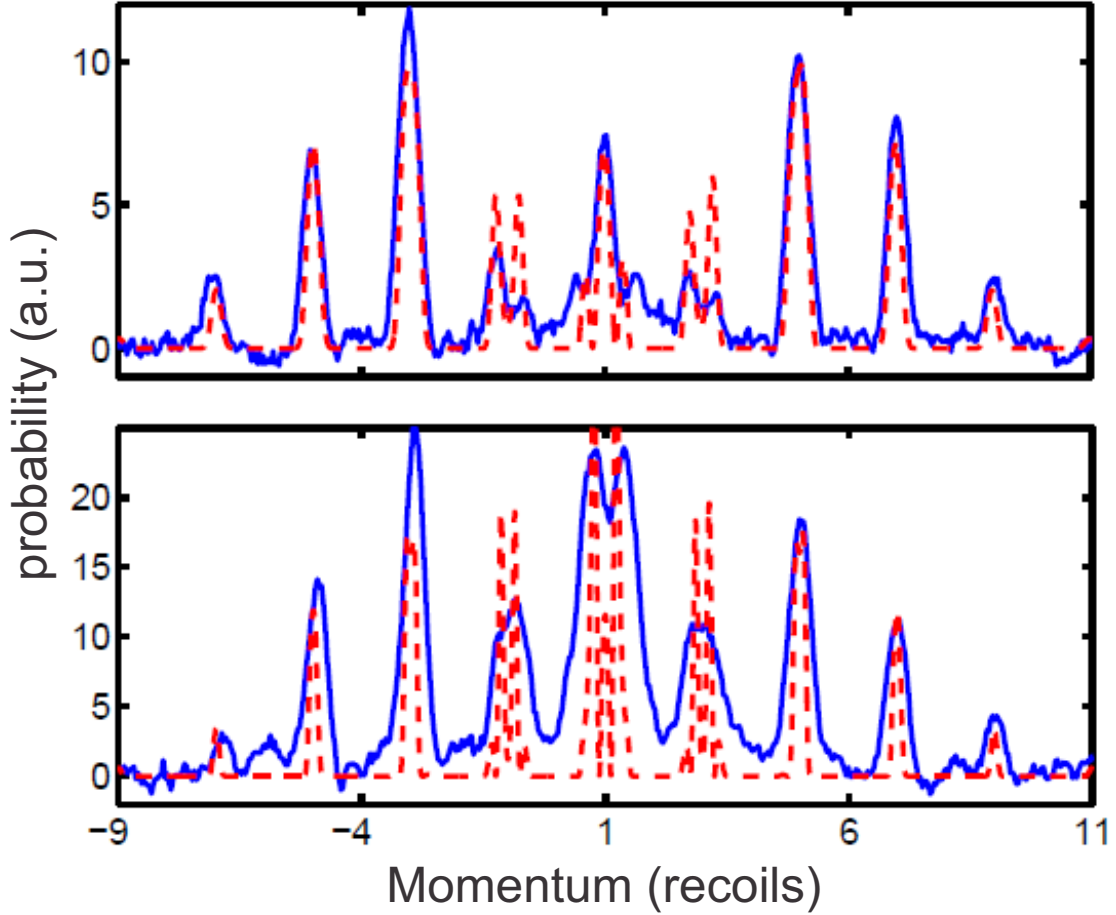}
       \caption{(Color online) The momentum distributions of the atoms after four kicks. The
momentum distributions for the kick period of $T=T_{T}/2$ (Top)
       and $T=T_{T}$ (bottom) are shown in the experiment
       (solid blue lines) and the simulation (red dashed lines) for $T=T_{T}/2$. The
initial momentum $p_{i}=p_{rec}$ in both cases and $\phi_{d}\approx 1$.}
       \label{figfive}
\end{figure}
\subsection{Details of the momentum distribution}
 In a similar experiment, the momentum distributions obtained as a result of
four kicks are shown in Fig.~\ref{figfive} for
 the antiresonance case (top) and the resonance
case (bottom). A change in widths of the momentum distributions for different diffraction orders is observed in both cases. The observed difference in widths can easily be related to the
separation of one velocity peak into multiple peaks in the simulation. For instance, the velocity peak at a momentum $p = 1$ recoil is split into three, with varying heights in the two
situations. This would be caused by the peak disappearing for $\Delta_{p} = 0.2$ recoils, and subsequently re-appearing for $\Delta_{p} = 0.4$. In the final momentum distribution, the latter
peaks appear even though they are not strongly weighted by the initial velocity distribution.

Similarly, the peaks at $p = 3$ and $p= -1$ recoils are split into two. This would be caused by the peaks appearing at a small $\Delta_{p}$. In addition,
 the peaks at $p = -(5, 7)$ recoils are found to be considerably narrower than the initial momentum distribution, which is caused by the fact that this diffraction peak only appears
for very small $\Delta_{p}$. These observations can be verified in Fig.~\ref{deltap} by drawing a horizontal line at fixed $\Delta_{p}$ and finding the diffraction orders. In the simulation, it
has been found that the width of resonances as a function of the initial momentum gets smaller in the presence of a large number of kicks.
\section{Conclusions}
 In this paper we present an experimental study of quantum resonances in the
$\delta-$kicked rotor system in detail. Quantum resonances are intrinsic to quantum mechanics and are therefore useful in studies related to the phenomenon of quantum chaos. We believe that the
results provided in this paper contribute to a deeper understanding of quantum dynamics. Using the kinetic energy measurements,
 quantum resonances and antiresonances have been observed for a range of kicks
in an AOKR experiment. In particular, we have shown how these resonances behave
as we increase
 the number of kicks. For large number of kicks, the energy of the atom is very
sensitive to each induced phase shift and is difficult to analyze accurately because of limited experimental resolution. Within the limits of experimental resolution we have found that these
resonances get sharper as the number of kicks is increased, in good agreement with simulations. Moreover, higher order resonances that arise from fractional quantum revivals have been observed
for pulse periods of (1/4)$T_{T}$ and (1/5)$T_{T}$ for five and ten kicks.

 The study of quantum resonant effects in this work has been further extended to
analyze the effect of the initial momentum of the atoms on quantum resonances in the AOKR. The first three quantum resonances have been examined for two and four kicks. A sinusoidal dependence
of the energy on the initial momentum for two kicks has been observed. By increasing the number of kicks, the system becomes extremely sensitive to the initial momentum of the atoms. A more
complex structure of the energy is observed for four kicks. For four kicks, the maxima in the experiment is in agreement with that in the simulation. The small-period oscillations in the
simulations, however, were not resolved in the experiment due to limited resolution for these settings. The maximum in the energy curves is observed to be much narrower than those for small
number of kicks, which is again in good agreement with simulations. With these experiments, it has been found that by applying a small frequency difference between the beams constituting the
standing wave, we can dial any initial velocity we choose for the atoms with respect to the standing wave. In general, we found that the energy is periodic with initial momentum.

  In the future it would be an interesting study to examine the dependence of the
initial velocity of the atoms on the fractional resonant effects in the AOKR. In the ``ballistic'' regime, i.e. for the kick period close to one of the ``quantum resonances'', a recent proposal
~\cite{McDowall2009} discusses the re-creation of the wave function by a strong, final pulse. The aim would be to confirm their predictions and investigate the influence of decoherence.  In
particular, it would be interesting to investigate the influence of the non-linearity caused by having a dense atom cloud. By manipulating the relative phase of the delta-kicks, a primitive
``quantum computer'' was recently demonstrated in ~\cite{Sadgrove2008}. The aim would be to take this idea further, making use of the excellent control we have of the kick laser field. As these
systems are very sensitive to initial conditions, the aim would be to investigate and measure these initial conditions, in particular to set an initial velocity and measure the recoil frequency,
which is proportional to $\hbar/m$, where $m$ is the mass of the particle. This was outlined in a recent study ~\cite{Horne2011}.

\acknowledgments{The authors acknowledge the University of Auckland Research Fund for financial support. The authors would also like to thank Fabienne Haupert for her work on the initial setup.
A. U. acknowledges the Higher Eduction Commission (HEC) of Pakistan for financial support.}


\end{document}